\documentclass[12pt,a4paper,final]{iopart}
\usepackage{iopams}
\usepackage{graphicx}
\usepackage[breaklinks=true,colorlinks=true,linkcolor=blue,urlcolor=blue,citecolor=blue]{hyperref}

\usepackage{centernot}
\usepackage{color}

\newcommand{\RR}{\mathbf{R}}
\newcommand{\rr}{\mathbf{r}}
\newcommand{\xx}{\mathbf{x}}
\newcommand{\yy}{\mathbf{y}}
\newcommand{\XX}{\mathbf{X}}
\newcommand{\YY}{\mathbf{Y}}
\newcommand{\kk}{\mathbf{k}}
\newcommand{\mm}{\mathrm{m}}

\newcommand{\CC}{\mathbf{C}}

\newcommand{\qq}{\mathbf{q}}

\newcommand{\eee}{\mathbf{e}}

\newcommand{\uu}{\mathbf{u}}
\newcommand{\dd}{\mathrm{d}}

\newcommand{\FF}{\mathcal{F}}
\newcommand{\ii}{{\mathrm{i}}}
\newcommand{\degen}{{\mathcal{D}}}

\newcommand{\be}{\begin{equation}}
\newcommand{\ee}{\end{equation}}
\newcommand{\bea}{\begin{eqnarray}}
\newcommand{\eea}{\end{eqnarray}}

\newcommand{\qed}{\nobreak \ifvmode \relax \else
      \ifdim\lastskip<1.5em \hskip-\lastskip
      \hskip1.5em plus0em minus0.5em \fi \nobreak
      \vrule height0.75em width0.5em depth0.25em\fi}

\begin{document}
\title[Ideal Fermi gas interaction-sensitive states and unitary Fermi gas virial expansion]
{The interaction-sensitive states of a trapped two-component ideal Fermi gas and application to the virial expansion of the unitary Fermi gas}

\author{Shimpei Endo and Yvan Castin$^*$}
\address{Laboratoire Kastler Brossel, ENS-PSL, CNRS, UPMC-Sorbonne Universit\'es and Coll\`ege de France, Paris, France}
\address{$^*$yvan.castin@lkb.ens.fr}

\begin{abstract}
We consider a two-component ideal Fermi gas in an isotropic harmonic potential. Some eigenstates have a wavefunction that vanishes 
when two distinguishable fermions are at the same location, and would be unaffected by $s$-wave contact interactions between the two components. 
We determine the other, interaction-sensitive eigenstates, using a Faddeev ansatz. This problem is nontrivial, due to degeneracies
and to the existence of unphysical Faddeev solutions. As an application we present a new conjecture 
for the fourth-order cluster or virial coefficient of the unitary Fermi gas, in good agreement with the numerical results of Blume and coworkers.
\end{abstract}


\section{Introduction and motivations}

We consider a three-dimensional trapped two-component ideal Fermi gas. The two components, noted as $\uparrow$ and $\downarrow$, correspond to two spin components of a single fermionic atomic species, or to two different fully polarised fermionic atomic species. The single particle masses $m_\uparrow$ and $m_\downarrow$ in each component may thus differ. There is no coherent coupling between the two states $\uparrow$ and $\downarrow$ so the total particle numbers in each component $N_\uparrow$ and $N_\downarrow$ are fixed, not simply the total particle number $N=N_\uparrow+N_\downarrow$. One can then take as reference spin configurations the $N_\uparrow+N_\downarrow$ configurations $\uparrow\ldots\uparrow \downarrow\ldots\downarrow$, where the wavefunction $\psi(\rr_1,\ldots,\rr_N)$ is antisymmetric under the exchange of the positions of the first $N_\uparrow$ particles, and under the exchange of the positions of the last $N_\downarrow$ particles.  The particles are trapped in the isotropic harmonic potential $U_\sigma(\rr)=m_\sigma\omega^2 r^2/2$ that depends on the component $\sigma=\uparrow,\downarrow$ in such a way that the angular oscillation frequency $\omega$ is $\sigma$-independent. In the experiments on cold atoms, where the interaction strength can be tuned via a Feshbach resonance \cite{Feshbach,InguscioGP}, our system is not a pure theoretical perspective and can be realised.

Imagine now that one turns on arbitrarily weak binary contact interactions between opposite spin particles. As the interaction has a zero range, it acts only among pairs of particles that approach in the $s$-wave. If one treats the interaction as a Dirac delta to first order in perturbation theory, some eigenstates of the ideal gas will experience an energy shift, some others will not. By definition, the shifted energy levels correspond to {\sl interaction-sensitive} states, and the unshifted ones to {\sl interaction-insensitive} states. This criterion can be implemented experimentally, by measuring the energy levels in the trap \cite{Jochim}.  Interestingly, the interaction-insensitive states have a vanishing wavefunction when any pair of particles converge to the same location; they thus remain unaffected by the interaction whatever its strength, provided that it remains zero range.

Even if this is an ideal gas problem, it is to our knowledge not treated in the classic literature. The interactions are usually of nonzero range, in nuclear physics or in quantum chemistry, and are not restricted to the $s$-wave channel; in this traditional context, our problem totally lacks physical motivation. This is probably why this problem was not mentioned in the classic book of Avery on hyperspherical harmonics \cite{Avery}, although the wavefunctions we are looking for are particular cases of hyperspherical harmonics, as we shall see.  Actually, specifically determining the interaction-sensitive states, and not simply all the eigenstates of trapped non-interacting fermions, is nontrivial due to the occurrence of large degeneracies of the unperturbed spectrum in an isotropic harmonic trap, so one faces the diagonalisation of large matrices in the degenerate perturbation theory, even if the problem can be first analytically reduced by the explicit construction of hyperspherical harmonics in Jacobi coordinates that are invariant (up to a global sign) under the exchange of identical fermions \cite{BlumeHS}. This degeneracy issue is reminiscent of the Fractional Quantum Hall Effect for contact interactions between cold atoms in an artificial magnetic field, where the macroscopic degeneracy of the Lowest Landau Level makes it nontrivial, even to first order perturbation theory, to determine the gapped phases induced by the interactions \cite{Jolicoeur}. The famous Laughlin wavefunction, when transposed to spinless bosons, is actually an interaction-insensitive state, which is thus automatically separated in energy space from the other, interaction-sensitive states when a repulsive contact interaction is turned on.  This is why, in reference \cite{Werner3corps}, the interaction-insensitive states were termed laughlinian states.

Another physical motivation is the calculation of the cluster or virial coefficients of the spatially homogeneous spin-$1/2$ unitary Fermi gas, where the opposite-spin fermions interact with a contact interaction of infinite $s$-wave scattering length. It is indeed now possible to measure the equation of state of the unitary gas with cold atoms \cite{virielENS,EOSTokyo,virielMIT}, from which one can extract the cluster coefficients up to fourth order \cite{virielENS,virielMIT}.  We recall that the cluster coefficients $b_{N_\uparrow, N_\downarrow}$ are, up to a factor, the coefficients of the expansion of the pressure of the thermal equilibrium gas of temperature $T$ in powers of the small fugacities $z_\sigma=\exp(\mu_\sigma/k_BT)$, that is in the low-density, non-degenerate limit where the chemical potential $\mu_\sigma$ of each spin component $\sigma$ tends to $-\infty$ \cite{Huang}. For the unitary gas, it is efficient to use the harmonic regulator technique of reference \cite{Comtet}, that is to determine the cluster coefficients $B_{N_\uparrow, N_\downarrow}(\omega)$ for the trapped system, in order to use its SO(2,1) dynamical symmetry \cite{Pitaevskii,CastinCRAS,WernerCastinSO21}; then one takes the $\omega\to 0$ limit to obtain the $b_{N_\uparrow,N_\downarrow}$.  It only remains to solve trapped few-body problems, since $B_{N_\uparrow, N_\downarrow}$ can be expressed in terms of the energy spectrum of all the $n_\uparrow+n_\downarrow$ systems, with $n_\sigma\leq N_\sigma$.  For the third cluster coefficient, this procedure was implemented numerically in reference \cite{Drummond}, and then analytically in reference \cite{Gao} by a generalisation to fermions of the inverse residue formula used for bosons in reference \cite{CastinCanada}. The predictions agree with the experimental results.  For the fourth virial coefficient, its numerical implementation by a direct calculation of the first few energy levels of four trapped fermions could not be pushed to low enough values of $\hbar \omega/k_B T$ to allow for a successful comparison with experiment \cite{Blumeb4}, and its analytical implementation is still an open problem \cite{pasEfim4}.

In all these calculations, what is actually computed is the difference $\Delta B_{N_\uparrow, N_\downarrow}(\omega)$ between the cluster coefficients of the unitary gas and of the ideal gas, so as to get rid of the contributions of the interaction-insensitive states, which are common to the two systems and exactly cancel. So for the ideal gas, one must determine the energy levels of the interaction-sensitive states.  For the $2+1$ fermionic systems (or equivalently for 3 bosons) this was done analytically in references \cite{Drummond,CastinCanada}. For the $3+1$ and $2+2$ fermionic systems, this was done numerically for the first few energy levels in reference \cite{Blumeb4}.  In this work, we obtain from a Faddeev ansatz an analytical prediction for all values of $N_\uparrow$ and $N_\downarrow$.  We then face a subtlety of the problem, that was already known for $2+1$ fermions \cite{Werner3corps}: some of the energy levels predicted by our Faddeev ansatz are unphysical and must be disregarded, since the corresponding wavefunction is zero.  We solve this issue for $3+1$ and $2+2$ fermions, with a general analytical reasoning completed for $2+2$ fermions by a case by case analysis.

The paper is organised as follows. In section \ref{sec:les_bases}, we introduce the basic theory tools already available in the literature \cite{WernerCastinLivre}, allowing us to reduce the problem to a zero energy free space problem with a wavefunction of the Faddeev form, each free space solution, characterised by a scaling exponent $s$, giving rise in the trapped system to a semi-infinite ladder of interaction-sensitive energy levels equispaced by $2\hbar\omega$.  In section \ref{sec:le_resultat_en_gros}, we give the corresponding scaling exponents $s$ for an arbitrary $N_\uparrow+N_\downarrow$ spin configuration.  In section \ref{sec:non_physique}, we investigate for $N=4$ the unphysical values of $s$, that are artifacts of the Faddeev ansatz.  In section \ref{sec:viriel} we present some applications to the cluster expansion of the unitary gas,
with a new conjecture for the fourth cluster coefficient and a comparison to the numerical results of \cite{Blumeb4}.  We conclude in section \ref{sec:conclusion}.

\section{The theoretical building blocks}

In this section, we remind the reader how, due to scale invariance, the energy levels of the trapped system can be deduced from the zero-energy free space solutions, more precisely from their scaling exponents (for a review, see reference \cite{WernerCastinLivre}).  We also explain, building on a footnote of reference \cite{CastinCanada}, how the interaction-sensitive states of the ideal gas can be singled out from the interaction-insensitive ones using a Faddeev ansatz for the $N$-body wavefunction.

\label{sec:les_bases}
\subsection{Scale invariance and the resulting SO(2,1) symmetry}
\label{subsec:SO21}

In free space, the ideal gas Hamiltonian
\be
H_{\rm free} = \sum_{i=1}^{N_\uparrow} -\frac{\hbar^2}{2m_\uparrow} \Delta_{\rr_i} +\sum_{i=N_\uparrow+1}^{N} -\frac{\hbar^2}{2m_\downarrow} \Delta_{\rr_i}
\label{eq:Hfree}
\ee
is scale invariant. Therefore, if $\psi_{\rm free}(\rr_1,\ldots,\rr_N)$ is an eigenstate of $H_{\rm free}$ with the eigenvalue zero,
\be
H_{\rm free} \psi_{\rm free}=0
\label{eq:Schrodi}
\ee
so is $\psi_{\rm free}^\lambda(\rr_1,\ldots,\rr_N)\equiv \psi_{\rm free}(\lambda \rr_1,\ldots,\lambda \rr_N)$, where all coordinates are multiplied by the same arbitrary scaling factor $\lambda>0$. An elementary consequence is that one can choose $\psi_{\rm free}$ to be scale invariant, which means that the wavefunctions $\psi_{\rm free}^\lambda$ and $\psi_{\rm free}$ are proportional. The corresponding {\sl scaling exponent} $s$ of $\psi_{\rm free}$ is then conveniently defined as follows:
\be
\psi_{\rm free}(\lambda \rr_1,\ldots, \lambda \rr_N) = \lambda^{s-\frac{3N-5}{2}} \psi_{\rm free}(\rr_1,\ldots,\rr_N) \ \ \ \ \forall \lambda>0
\label{eq:defs}
\ee
In other words, $\psi_{\rm free}(\rr_1,\ldots,\rr_N)$ is a positively homogeneous function of the coordinates of degree $s-(3N-5)/2$.  Further using the free space translational invariance, one imposes that the centre of mass of the system is at rest:
\be
\psi_{\rm free}(\rr_1+\uu,\ldots,\rr_N+\uu)=\psi_{\rm free}(\rr_1,\ldots,\rr_N) \ \ \ \forall \uu \in \mathbb{R}^3
\label{eq:invartrans}
\ee

A more elaborate consequence is that one can generate from $\psi_{\rm free}$ a semi-infinite ladder of exact eigenstates of the Hamiltonian $H$ of the {\sl trapped} system,
\be
H=H_{\rm free}+H_{\rm trap},  \ \ \ H_{\rm trap}=\sum_{i=1}^{N_\uparrow} \frac{1}{2}m_\uparrow \omega^2 r_i^2 + \sum_{i=N_\uparrow+1}^{N} 
 \frac{1}{2}m_\downarrow \omega^2 r_i^2
\ee
Each rung of the ladder is indexed by a quantum number $q\in \mathbb{N}$. The corresponding unnormalised wavefunction is \cite{WernerCastinSO21}
\be
\psi_q(\rr_1,\ldots,\rr_N)=L_q^{(s)}(R^2/a_{\rm ho}^2) e^{-\sum_{i=1}^{N} m_i \omega r_i^2/(2\hbar)} \psi_{\rm free}(\rr_1,\ldots,\rr_N)
\label{eq:mapping}
\ee
where $m_i$ is the mass of particle $i$, $R$ is the internal hyperradius of the $N$ particles
\be
R \equiv \left[\frac{1}{m_u}\sum_{i=1}^{N} m_i (\rr_i-\CC)^2\right]^{1/2}
\label{eq:defR}
\ee
involving the position of the centre of mass $\CC=\left(\sum_{i=1}^{N} m_i\rr_i\right)/\left(\sum_{i=1}^{N} m_i\right)$ of the system and some arbitrary mass reference $m_u$,
$a_{\rm ho}=[\hbar/(m_u\omega)]^{1/2}$ is the corresponding harmonic oscillator length and $L_q^{(s)}(X)$ is the generalised Laguerre polynomial of degree $q$:
\be
L_q^{(s)}(X)\equiv \frac{X^{-s} e^X}{q!} \frac{\dd^q}{\dd X^q} (X^{q+s}e^{-X})
\ee
In a harmonic potential, the centre of mass motion and the relative motion are separable.  Since the wavefunction $\psi_{\rm free}$ and the internal variable $R$ are translationally invariant, the wavefunction $\psi_q$ corresponds to the centre of mass motion in its ground state with an energy $3\hbar\omega/2$. The eigenenergy of $\psi_q$ is thus $E_q=\frac{3}{2}\hbar\omega + E_q^{\rm rel}$, where $E_q^{\rm rel}$ is the relative or internal eigenenergy, given by \cite{WernerCastinSO21}
\be
E_q^{\rm rel}=(s+1+2q) \hbar \omega \ \ \forall q \in \mathbb{N}
\label{eq:Eq}
\ee
Physically, this ladder structure reflects the fact that scale invariant systems acquire in a harmonic trap an exact breathing mode of angular frequency $2\omega$ \cite{Pitaevskii,CastinCRAS}. This mode, when quantised, is a bosonic mode of Hamiltonian $2\hbar\omega \hat{b}^\dagger \hat{b}$, where the creation operator $\hat{b}^\dagger$ and the annihilation operator $\hat{b}$, obeying the usual commutation relation $[\hat{b},\hat{b}^\dagger]=1$, are raising and lowering operators in each semi-infinite ladder, exciting and deexciting the breathing mode by one quantum \cite{WernerCastinSO21}.  Mathematically, this reflects the SO(2,1) dynamical symmetry of the trapped system, $H$ being part of a SO(2,1) Lie algebra.

One can show that the mapping (\ref{eq:mapping}) is complete, meaning that all eigenstates in the trap with a ground state centre of mass are obtained if one uses all possible $\psi_{\rm free}$ \cite{WernerCastinSO21}.  The trapped problem is thus reduced to a zero energy free space problem in the rest frame and we only need in practice to determine the scaling exponents $s$ of the corresponding interaction-sensitive eigenstates $\psi_{\rm free}$.

\subsection{The Faddeev ansatz in real space and in Fourier space}

To filter out the interaction-sensitive states of the ideal gas, we use the technique proposed in a footnote of reference \cite{CastinCanada}.  We introduce a zero-range interaction between the opposite spin fermions, with a finite $s$-wave scattering length $a$, in the form of Wigner-Bethe-Peierls contact conditions on the $N$-body wavefunction \cite{Wigner,BethePeierls}: for all $\uparrow\downarrow$ pairs, that is for all particle indices $i$ and $j$, with $1\leq i\leq N_\uparrow$ and $N_\uparrow+1\leq j\leq N$, there exists a function $A_{ij}$, called the {\sl regular part},
such that
\be
\psi_{\rm free}(\rr_1,\ldots,\rr_N) \stackrel{r_{ij}\to 0}{=} \left(\frac{1}{r_{ij}}-\frac{1}{a}\right) A_{ij}((\rr_k-\RR_{ij})_{k\neq i,j})+O(r_{ij})
\label{eq:WBP}
\ee
Here, the relative coordinates $\rr_{ij}=\rr_i-\rr_j$ of particles $i$ and $j$ tend to zero at a fixed position $\RR_{ij}=(m_i \rr_i+m_j\rr_j)/(m_i+m_j)$ of their centre of mass, different from the positions $\rr_k$, $1\leq k\leq N$ and $k\neq i,j$, of the other particles. Due to the assumed translational invariance (\ref{eq:invartrans}) of the wavefunction in free space, we have directly considered here $A_{ij}$ as a function of the relative positions $\rr_k-\RR_{ij}$.  The idea now is that the interaction-insensitive states have identically zero regular parts, $A_{ij}\equiv 0$, for all $i$ and $j$. The interacting states, on the contrary, have nonzero regular parts, and they converge, when $a\to 0$, to the desired interaction-sensitive states of the ideal gas.

To solve Schr\"odinger's equation in the presence of the contact conditions (\ref{eq:WBP}), one formulates it in the framework of distributions \cite{HouchesCastin,Petrov}.  Due to the $1/r_{ij}$ singularities, to the identity $\Delta_{\rr}(1/r)=-4\pi\delta(\rr)$ and to the rewriting
\be
-\frac{\hbar^2}{2m_i}\Delta_{\rr_i} - \frac{\hbar^2}{2m_j}\Delta_{\rr_j} = -\frac{\hbar^2}{2M_{\uparrow\downarrow}}\Delta_{\RR_{ij}}
-\frac{\hbar^2}{2\mu_{\uparrow\downarrow}}\Delta_{\rr_{ij}}
\label{eq:cinij}
\ee
with $M_{\uparrow\downarrow}=m_\uparrow+m_\downarrow$ the total mass and $\mu_{\uparrow\downarrow}=m_\uparrow m_\downarrow/M_{\uparrow\downarrow}$ the reduced mass of two opposite spin particles, equation~(\ref{eq:Schrodi}) acquires three-dimensional Dirac delta terms in the right-hand side:
\be
H_{\rm free} \psi_{\rm free}(\rr_1,\ldots, \rr_N)= \sum_{i=1}^{N_\uparrow} \sum_{j=N_\uparrow+1}^{N} \frac{2\pi\hbar^2}{\mu_{\uparrow\downarrow}} 
A_{ij}((\rr_k-\RR_{ij})_{k\neq i,j}) \delta(\rr_{ij})
\label{eq:Schrodi2}
\ee
Multiplying formally equation~(\ref{eq:Schrodi2}) by the inverse of the operator $H_{\rm free}$, that is expressing its solution in terms of the Green's function of a $3N$ dimensional Laplacian, we obtain $\psi_{\rm free}$ as a sum over $i$ and $j$ of Faddeev components, $\displaystyle\psi_{\rm free}=\sum_{i=1}^{N_\uparrow} \sum_{j=N_\uparrow+1}^{N} \FF_{ij}$, with
\be
\FF_{ij} \equiv \frac{1}{H_{\rm free}} \frac{2\pi\hbar^2}{\mu_{\uparrow\downarrow}} A_{ij}((\rr_k-\RR_{ij})_{k\neq i,j}) \delta(\rr_{ij})
\ee

Let us review the symmetry properties of the Faddeev components. First, the $(i,j)$ source term in equation (\ref{eq:Schrodi2}) is translationally invariant, as well as $H_{\rm free}$, and so is $\FF_{ij}$. Second, the $(i,j)$ source term is invariant by rotation of $\rr_{ij}$ at fixed $\RR_{ij}$, and so is $\FF_{ij}$ because the $i$ and $j$ Laplacians in $H_{\rm free}$ can be rewritten as in equation (\ref{eq:cinij}); as a consequence, $\FF_{ij}$ depends on $\rr_{ij}$ only through its modulus $r_{ij}$. Third, due to the fermionic exchange symmetry, the regular parts $A_{ij}$ are not functionally independent and coincide with $A_{1N_\uparrow+1}$ up to a sign, which is the signature of the permutation that maps  $(1,\ldots,i,\ldots,N_\uparrow,N_\uparrow+1,\ldots,j,\ldots,N)$ to $(i,1,\ldots,i-1,i+1,\ldots,N_\uparrow,j,N_\uparrow+1,\ldots,j-1,j+1,\ldots,N)$:
\be
A_{ij}((\xx_k)_{k\neq i,j})=(-1)^{i-1}(-1)^{j-(N_\uparrow+1)} A_{1N_\uparrow+1}((\xx_k)_{k\neq i,j})
\ee
Similarly, the Faddeev components can all be expressed in terms of the first Faddeev component $\FF_{1N_\uparrow+1}$, noted as $\FF$ for concision.  Fourth, at fixed $(i,j)=(1,N_\uparrow+1)$, the fermionic exchange symmetry among the last $N_\uparrow-1$ spin $\uparrow$ particles and among the last $N_\downarrow-1$ spin $\downarrow$ particles imposes that $\FF(r;(\xx_k)_{k\neq 1,N_\uparrow+1})$ is a fermionic function of its first $N_\uparrow-1$ vectorial variables, and a fermionic function of its last $N_\downarrow-1$ vectorial variables:
\bea
\FF(r;(\xx_{\sigma(k)})_{2\leq k\leq N_\uparrow}, (\xx_k)_{N_\uparrow+2\leq k\leq  N}) &=& \epsilon(\sigma) \FF(r;(\xx_k)_{k\neq 1, N_\uparrow+1}) \\
\FF(r;(\xx_k)_{2\leq k\leq N_\uparrow}, (\xx_{\sigma(k)})_{N_\uparrow+2\leq k\leq  N}) &=& \epsilon(\sigma) \FF(r;(\xx_k)_{k\neq 1, N_\uparrow+1})
\eea
where $\sigma$, of signature $\epsilon(\sigma)$, is any permutation of $N_\uparrow-1$ or of $N_\downarrow-1$ objects, respectively.

We finally take the non-interacting limit $a\to 0$ and we obtain the following Faddeev ansatz for the wavefunction of the interaction-sensitive states of the ideal gas:
\be
\psi_{\rm free}(\rr_1,\ldots,\rr_N) = \sum_{i=1}^{N_\uparrow} \sum_{j=N_\uparrow+1}^{N} (-1)^{i-1+j-(N_\uparrow+1)}
\FF(r_{ij}; (\rr_k-\RR_{ij})_{k\neq i,j})
\label{eq:ansatz_de_Faddeev}
\ee
The key point is that, in the $(i,j)$ component, particles $i$ and $j$ approach in a purely $s$-wave relative motion, which is a necessary condition for them to be sensitive to $s$-wave contact interactions. 

It will be shown in section \ref{sec:non_physique} that this is not always sufficient to make $\psi_{\rm free}$ interaction-sensitive, because the Faddeev ansatz leads in some cases to $\psi_{\rm free}\equiv 0$, that is to unphysical solutions.  To investigate this point, the momentum space version of the Faddeev ansatz will be helpful. It was originally put forward for the interacting gas \cite{pasEfim4,Petrov,CMP,Cras4corps}, but it can also be used to find the unphysical solutions of the non-interacting gas, as they are the same at all interaction strength.  Introducing the Fourier representation of the regular part, 
\be
\frac{2\pi \hbar^2}{\mu_{\uparrow\downarrow}} A_{1N_\uparrow+1}((\xx_k)_{k\in I})= \frac{1}{(2\pi)^3}\int \prod_{j\in I} \frac{{\dd}^3k_j}{(2\pi)^3} D((\kk_j)_{j\in I})
e^{\ii\sum_{j\in I} \kk_j \cdot \xx_j}
\ee
where all indices run over the set $I$ of integers from $2$ to $N$ different from $N_\uparrow+1$,
\be
I=\{1,\ldots,N\} \setminus \{1,N_\uparrow+1\}
\label{eq:defI}
\ee
we obtain the Fourier space representation of the Faddeev component
\be
\FF(r;(\xx_k)_{k\in I})= \int \frac{{\dd}^3q}{(2\pi)^6} \prod_{j\in I} \frac{{\dd}^3k_j}{(2\pi)^3} \frac{D((\kk_j)_{j\in I}) e^{\ii\sum_{j\in I} \kk_j \cdot \xx_j}
e^{\ii\qq\cdot\rr}}{\frac{\hbar^2 q^2}{2\mu_{\uparrow\downarrow}}+\frac{\hbar^2(\sum_{j\in I}\kk_j)^2}{2M_{\uparrow\downarrow}}+\sum_{j\in I} \frac{\hbar^2k_j^2}{2m_j}}
\label{eq:compFadFour}
\ee
where $\rr$ is any vector of modulus $r$ and, physically, $\qq$ is the relative wave vector of particles $1$ and $N_\uparrow+1$ and $-\sum_{j\in I}\kk_j$ their total wave vector. This corresponds to the following ansatz for the Fourier transform of the $N$-body wavefunction:
\be
\tilde{\psi}_{\rm free}(\kk_1,\ldots,\kk_N)= \frac{\delta(\sum_{i=1}^{N} \kk_i)}{\sum_{i=1}^{N} \frac{\hbar^2 k_i^2}{2m_i}}
\sum_{i=1}^{N_\uparrow}\!\sum_{j=N_\uparrow+1}^{N} (-1)^{i-1+j-(N_\uparrow+1)} D((\kk_n)_{n\neq i,j})
\label{eq:ansatz_psi_tilde}
\ee
in agreement with reference \cite{Ludovic}. Obviously, $D((\kk_j)_{j\in I})$ is fermionic with respect to its first $N_\uparrow-1$ vectorial variables, and fermionic with respect to its last $N_{\downarrow}-1$ vectorial variables, exactly as $A_{1 N_{\uparrow}+1}$ and $\FF$. Also, its scaling exponent in the unitary limit can be expressed in terms of the scaling exponent $s$ of the wavefunction through the usual power-counting argument for the Fourier transform:
\be
D(\lambda (\kk_n)_{n\in I})= \lambda^{-(s+\frac{3N-5}{2})} D((\kk_n)_{n\in I})\ \ \ \forall \lambda>0
\label{eq:scaD}
\ee

\section{Scaling exponents of the interaction-sensitive states of the ideal gas}
\label{sec:le_resultat_en_gros}
\subsection{The general result for arbitrary particle numbers}

It is well known from the one-body case that all eigenstates of the trapped system Hamiltonian $H$ are products of polynomials in the $3N$ coordinates of the particles and of the Gaussian factor appearing in equation~(\ref{eq:mapping}), and so are the $\psi_q$. Taking $q=0$ in that equation, so that $L_q^{(s)}\equiv 1$, one sees that the free space eigenstate $\psi_{\rm free}(\rr_1,\ldots,\rr_N)$ is necessarily such a polynomial, and so is the Faddeev component $\FF$\footnote{Up to an appropriate coordinate rescaling to account for a possible mass difference $m_\uparrow\neq m_\downarrow$, $\psi_{\rm free}(\rr_1,\ldots,\rr_N)$ is a harmonic polynomial of degree $d$, since it is homogeneous and of zero Laplacian, and is translationally invariant, so it can be written as $R^d Y_d(\Omega)$, where $R$ is the internal hyperradius (\ref{eq:defR}), $\Omega$ is a set of hyperangles and $Y_d$ is a so-called hyperspherical harmonic. We are however only interested in the specific case of interaction-sensitive states, not discussed in the extensive book of Avery on hyperspherical harmonics \cite{Avery}.  The reference \cite{BlumeHS} implemented the formalism of Avery with cleverly chosen Jacobi coordinates ${\boldsymbol\rho}_i$, that are invariant (up to a global sign) under the exchange of identical fermions. For example, for equal mass $\uparrow\uparrow\downarrow$ fermions, it took ${\boldsymbol\rho}_1=\rr_1-\rr_2$ and ${\boldsymbol\rho}_2=(\rr_1+\rr_2)/2-\rr_3$.  To express however the fact that, in an interaction-sensitive state, the opposite-spin particles $1$ and $3$ approach in the $s$-wave, one must rather use a system of coordinates containing $\rr_{13}=\rr_1-\rr_3$, which is not invariant by permutation of particles $1$ and $2$.  This is why we introduced the extra ingredient of the Faddeev ansatz in equation~(\ref{eq:ansatz_de_Faddeev}), not relying on a specific choice of Jacobi coordinates.}.  As $\FF$ depends on the modulus $r$ (and not on the direction) of the relative coordinates of two $\uparrow\downarrow$ particles, only even powers of
$r$ can contribute to its expansion, hence the specific ansatz:
\be
\FF(r;(\xx_i)_{i\in I})=\sum_{k\geq 0} r^{2k} P_k((\xx_{i})_{i\in I})
\label{eq:devr2}
\ee
where the set $I$ is given by equation~(\ref{eq:defI}). Since $\psi_{\rm free}$ has a well defined scaling exponent $s$, see equation (\ref{eq:defs}), $\FF$ is a homogeneous polynomial of degree
\be
d = s -\frac{3N-5}{2}
\label{eq:ds}
\ee
so that each polynomial $P_k$ is homogeneous of degree $d-2k$ as long as $d-2k \geq 0$, otherwise it is identically zero and the series (\ref{eq:devr2}) terminates.  Last, $\psi_{\rm free}$ has a zero eigenenergy with respect to the free space Hamiltonian, see equation (\ref{eq:Schrodi}):
\be
H_{\rm free} \FF(r_{1N_{\uparrow}+1}; (\rr_i-\RR_{1 N_\uparrow+1})_{i\in I}) =0
\ee
From the explicit form (\ref{eq:Hfree}) of $H_{\rm free}$, modified with equation~(\ref{eq:cinij}) for the first $\uparrow$ and $\downarrow$ particles, and the chain rule of differential calculus, this is turned into a differential equation for
$\FF$:
\be
(\Delta_{\rr} + \hat{D})\FF(r;(\xx_i)_{i\in I})=0
\label{eq:laplaF}
\ee
Here $\Delta_{\rr}$, the usual three-dimensional Laplacian, can be restricted to its radial part $r^{-1}\partial_r^2(r\,\cdot)$ as far as the variable $r$ is concerned, and the differential operator $\hat{D}$, acting only on the vectorial variables of the Faddeev component, is given by
\be
\hat{D}=(1-t^2) \sum_{i=2}^{N_\uparrow} \Delta_{\xx_i} + t(2-t)\!\!\sum_{j=N_\uparrow+2}^{N} \Delta_{\xx_j} 
+2t(1-t) \sum_{i=2}^{N_\uparrow} \sum_{j=N_\uparrow+2}^{N}  \nabla_{\xx_i}\cdot \nabla_{\xx_j}
\ee
with the mass ratio
\be
t=\frac{m_\uparrow}{m_\uparrow+m_\downarrow} \in ]0,1[
\ee
When applied to the expansion (\ref{eq:devr2}), the equation (\ref{eq:laplaF}) gives a recurrence relation on the polynomials $P_k$,
\be
P_{k+1}((\xx_{i})_{i\in I}) = -\frac{1}{(2k+2)(2k+3)} \hat{D} P_k((\xx_{i})_{i\in I}) \ \ \forall k\geq 0
\ee
that ultimately allows to express them in terms of repeated actions of $\hat{D}$ on the polynomial $P_0$, the {\sl generating} polynomial.

In conclusion, to generate an arbitrary interaction-sensitive state $\psi_{\rm free}$ of zero energy in free space, one simply has to arbitrarily choose a polynomial $P_0((\xx_{i})_{i\in I})$ which is homogeneous of degree $d\in \mathbb{N}$ and antisymmetric under the exchange of its first $N_\uparrow-1$ vectorial variables and under the exchange of its last $N_\downarrow-1$ ones. The corresponding scaling exponent is given by (\ref{eq:ds}), and the corresponding Faddeev component is given by
\be
\FF(r;(\xx_{i})_{i\in I}) =\sum_{k\geq 0} \frac{r^{2k}(-\hat{D})^k}{(2k+1)!} P_0((\xx_{i})_{i\in I})
\label{eq:fadfin}
\ee
Then, one reconstructs the wavefunction $\psi_{\rm free}$ from equation (\ref{eq:ansatz_de_Faddeev}), and one generates a semi-infinite ladder of interaction-sensitive eigenstates of the trapped system using the mapping (\ref{eq:mapping}).

A natural choice, inspired by the rotational invariance, is to take as a basis of the polynomials of a single vectorial variable $\xx$ the set of homogeneous monomials 
\be
\xx\mapsto x^{2n+\ell} Y_\ell^{\mm}(\hat{\xx})
\label{eq:mono}
\ee
where $n \in \mathbb{N}$, $\hat{\xx}=\xx/x$ is the direction of $\xx$, parametrised by a polar angle and an azimuthal angle in spherical coordinates, and $Y_\ell^{\mm}$ is the corresponding spherical harmonic of orbital quantum number $\ell\in \mathbb{N}$ and azimuthal quantum number $\mm$ (in roman style to avoid confusion with a mass).  To construct $P_0$, one then puts one $\uparrow$ fermion in each $(n_i,\ell_i,\mm_i)$ state for $2\leq i\leq N_\uparrow$, and one $\downarrow$ fermion in each $(n_i,\ell_i,\mm_i)$ state for $N_\uparrow+2\leq i\leq N$, where the monomial states are chosen freely, except for the constraint that, within each spin manifold, they must be different and sorted in alphanumeric order to avoid multiple counting.  This simple construction leads to a total degree $d=\sum_{i\in I} (2n_i+\ell_i)$ and to a scaling exponent
\be
s=\frac{3N-5}{2} + \sum_{i\in I} (2n_i+\ell_i)
\label{eq:sgen}
\ee
According to the equation (\ref{eq:Eq}) the corresponding semi-infinite ladder of internal energies of interaction-sensitive states is
\be
E_q^{\rm rel}=\left(2q+\frac{3}{2}\right)\hbar\omega + \sum_{i\in I} \left(2n_i+\ell_i+\frac{3}{2}\right)\hbar\omega
\label{eq:jolie_forme}
\ee
This writing lends itself to a simple physical interpretation. The first term is an energy level of a harmonically trapped fictitious particle with zero angular momentum; this fictitious particle corresponds to the relative motion of two opposite spin fermions in the trap, and its restriction to the zero angular momentum sector ensures that it is sensitive to $s$-wave interactions. The second contribution in equation (\ref{eq:jolie_forme}) is any energy level of an ideal gas of $N_\uparrow-1$ spin $\uparrow$ fermions and $N_\downarrow-1$ spin $\downarrow$ fermions in the trap.  As we shall see, the result (\ref{eq:sgen}) has to be refined for $N>2$, as well as the transparent form (\ref{eq:jolie_forme}): some scaling exponents are unphysical and must be disregarded.

\subsection{Explicit results for $2+1$, $3+1$ and $2+2$ fermions}

For few-body systems, it is most convenient to take generating polynomials $P_0$ with a well defined total angular momentum $\ell$.  As the $r$ variable in equation (\ref{eq:fadfin}) carries a zero total angular momentum, the Faddeev component $\FF$ and the corresponding wavefunction $\psi_{\rm free}$ have a total angular momentum $\ell$. This conclusion extends to the trapped eigenstates $\psi_q$ since the variable $R$ and the Gaussian factor in equation (\ref{eq:mapping}) are rotationally invariant (remember that the centre of mass of the gas is in its ground state).  Similarly, the eigenstates have the same parity as $P_0$.

For $2+1$ fermions, the sum in equation (\ref{eq:sgen}) contains a single term. The generating exponents of the interaction-sensitive states are thus
\be
s_{\ell,n} = 2n+\ell+2, \ \ \ \forall (n,\ell) \in \mathbb{N}^2
\label{eq:s2p1}
\ee
with a degeneracy $2\ell+1$ and a parity equal to the natural parity $(-1)^\ell$. This agrees with reference \cite{Drummond}.

For $3+1$ fermions, the sum in equation (\ref{eq:sgen}) runs over the set $I=\{2,3\}$ so it involves the principal $n_i$ and orbital $\ell_i$ quantum numbers of particles $2$ and $3$.  As these are identical fermions, it is more convenient to use the  principal $(n_{\rm com},n_{\rm rel})\in\mathbb{N}^2$ and orbital $(\ell_{\rm com},\ell_{\rm rel})\in\mathbb{N}^2$ quantum numbers of their centre of mass and relative motions,  rewriting (\ref{eq:sgen}) as
\be
s=\frac{7}{2} + 2(n_{\rm com}+n_{\rm rel}) + \ell_{\rm com} +\ell_{\rm rel}
\label{eq:s31relcom}
\ee
and restricting to odd values of $\ell_{\rm rel}$.  From the composition of the two angular momenta $\ell_{\rm com}$ and $\ell_{\rm rel}$, an angular momentum $\ell$ can be obtained if and only if $(\ell_{\rm rel}, \ell_{\rm com},\ell)$ can be the lengths of the sides of a triangle, that is $|\ell_{\rm rel}-\ell_{\rm com}|\leq \ell \leq \ell_{\rm rel}+\ell_{\rm com}$, or more conveniently
\be
|\ell_{\rm rel}-\ell| \leq \ell_{\rm com} \leq \ell_{\rm rel}+\ell
\ee
The resulting parity $(-1)^{\ell_{\rm rel}+\ell_{\rm com}}$ can now differ from the natural parity $(-1)^\ell$. We write it as $\sigma (-1)^\ell$, where $\sigma=\pm 1$. Equivalently, $\ell_{\rm rel}+\ell_{\rm com}\equiv\ell+(1-\sigma)/2\ (\mbox{mod}\ 2)$ so we set
\be
s_{\ell,n}^{(\sigma)} = 2n + \ell + \frac{1-\sigma}{2} +\frac{7}{2},\ \ \ \ \forall (n,\ell)\in \mathbb{N}^2, \forall \sigma\in\{-1,1\}
\label{eq:s4corps}
\ee
It remains to sum the natural degeneracy $2\ell+1$ over all values of $(n_{\rm com},n_{\rm rel})$ and $(\ell_{\rm com},\ell_{\rm rel})$ to obtain the full degeneracy 
\be
\degen_{\ell,n}^{(\sigma)}= (2\ell+1) \sum_{\ell_{\rm rel}\in2\mathbb{N}+1} \sum_{\ell_{\rm com}=|\ell-\ell_{\rm rel}|}^{\ell+\ell_{\rm rel}}
\sum_{(n_{\rm rel},n_{\rm com})\in \mathbb{N}^2} \delta_{2(n_{\rm rel}+n_{\rm com}),p-\ell_{\rm rel}-\ell_{\rm com}}
\ee
where $\delta$ is the Kronecker symbol and $p=\ell+2n+\frac{1-\sigma}{2}$.  The sum over $(n_{\rm rel},n_{\rm com})$ is readily performed using the variables $n_{\rm tot}=n_{\rm rel}+n_{\rm com}\in \mathbb{N}$ and $n_{\rm rel}$ ranging from $0$ to $n_{\rm tot}$, as the summand depends only on $n_{\rm tot}$.  This sum is nonzero only if $\ell_{\rm com}\leq p-\ell_{\rm rel}$ and if $p-\ell_{\rm rel}-\ell_{\rm com}$ is even, this second condition being taken care of by inclusion of a factor $[1+(-1)^{p-\ell_{\rm rel}-\ell_{\rm com}}]/2$. Similarly, one introduces a factor $[1-(-1)^{\ell_{\rm rel}}]/2$ to take care of the oddness of $\ell_{\rm rel}$ due to the fermionic antisymmetry. This leads to
\bea
\degen_{\ell,n}^{(\sigma)} &=& (2\ell+1) \sum_{\ell_{\rm rel}\in\mathbb{N}} 
\sum_{\ell_{\rm com}=|\ell-\ell_{\rm rel}|}^{\mathrm{min}\, (\ell+\ell_{\rm rel},p-\ell_{\rm rel})} 
\left[\frac{1-(-1)^{\ell_{\rm rel}}}{2}\right]\, \left[\frac{1+(-1)^{p-\ell_{\rm rel}-\ell_{\rm com}}}{2}\right] \nonumber \\
&&\times \left(1+\frac{p-\ell_{\rm rel}-\ell_{\rm com}}{2}\right)
\label{eq:dinterm}
\eea
The double sum is calculated by distinguishing the cases $0\leq\frac{p-\ell}{2}\leq\ell$ and $\ell<\frac{p-\ell}{2}$ \footnote{In the first case, $\ell_{\rm com}$ runs from $\ell-\ell_{\rm rel}$ to $\ell+\ell_{\rm rel}$ for $0\leq \ell_{\rm rel} \leq \frac{p-\ell}{2}$, from $\ell-\ell_{\rm rel}$ to $p-\ell_{\rm rel}$ for $\frac{p-\ell}{2} < \ell_{\rm rel}\leq \ell$, and from $\ell_{\rm rel}-\ell$ to $p-\ell_{\rm rel}$ for $\ell < \ell_{\rm rel} \leq \frac{p+\ell}{2}$. In the second case, $\ell_{\rm com}$ runs from $\ell-\ell_{\rm rel}$ to $\ell+\ell_{\rm rel}$ for $0\leq \ell_{\rm rel} \leq \ell$, from $\ell_{\rm rel}-\ell$ to $\ell+\ell_{\rm rel}$ for $\ell < \ell_{\rm rel}\leq \frac{p-\ell}{2}$, and from $\ell_{\rm rel}-\ell$ to $p-\ell_{\rm rel}$ for $\frac{p-\ell}{2} < \ell_{\rm rel} \leq \frac{p+\ell}{2}$. In both cases, the sum over $\ell_{\rm com}$ is empty for $\ell_{\rm rel}> \frac{p+\ell}{2}$.} and the subcases $\sigma=\pm 1$. We finally obtain for $3+1$ fermions:
\be
\degen_{\ell,n}^{(\sigma)}=\frac{(2\ell\!+\!1)\!\left\{(2\ell\!+\!1\!+\!\sigma)(n\!+\!1)(n\!+\!2)\!-\![\sigma\!+\!(-1)^\ell]\left[n\!+\!1\!+\!\frac{1+(-1)^n}{2}\right]\right\}}{8}
\label{eq:d3p1}
\ee

For $2+2$ fermions, the particles indexed by the set $I=\{2,4\}$ are now distinguishable.  One reuses the last calculation, simply relaxing the parity condition on $\ell_{\rm rel}$, that is removing the factor $\frac{1-(-1)^{\ell_{\rm rel}}}{2}$ in equation (\ref{eq:dinterm}). The scaling exponent of the interaction-sensitive states, written as in equation (\ref{eq:s4corps}), now has a degeneracy
\be
\degen_{\ell,n}^{(\sigma)}=(2\ell+1)\left(\ell+\frac{1+\sigma}{2}\right) \frac{(n+1)(n+2)}{2}
\label{eq:d2p2}
\ee
Both results (\ref{eq:d3p1}) and (\ref{eq:d2p2}) vanish at all $n$ for $(\ell,\sigma)=(0,-1)$ as they should, since isotropic states of two particles (corresponding to the set $I$) necessarily have the natural parity $+1$.  Both also include unphysical scaling exponents corresponding to a vanishing Faddeev ansatz wavefunction (\ref{eq:ansatz_de_Faddeev}); this will be corrected in section \ref{sec:non_physique}.

\section{Refining the theory: exclusion of the unphysical solutions} 
\label{sec:non_physique}

For $N>2$, some of the scaling exponents predicted in section \ref{sec:le_resultat_en_gros} are unphysical, as they do not correspond to any interaction-sensitive state of the ideal gas:  the corresponding Faddeev ansatz wavefunction (\ref{eq:ansatz_de_Faddeev}) vanishes, due to the destructive interference of its individually nonzero Faddeev components. This problem was already solved for $N=3$: there is a single unphysical solution \cite{Werner3corps}, corresponding to $(n,\ell)=(0,0)$ in equation (\ref{eq:s2p1}), that is to a generating polynomial $P_0=1$ and a Faddeev component $\FF=1$ obviously giving $\psi_{\rm free}\equiv 0$ in equation (\ref{eq:ansatz_de_Faddeev}). To our knowledge, it is still open for $N>3$.  We investigate it explicitly for $N=4$. An infinite number of unphysical solutions is easily predicted by a formal reasoning in Fourier space with divergent integrals, in section \ref{subsec:Fourier}.  Then we perform a real space calculation on a case by case basis in section \ref{subsec:espace_reel}: for a specific unphysical solution, taken as an example, we confirm the value of the generating polynomial $P_0$ predicted by the general Fourier space reasoning, giving a meaning to the divergent integrals by analytic continuation; we also show that some unphysical solutions are missed by the Fourier space reasoning.

\subsection{Reasoning in Fourier space for $N=4$}
\label{subsec:Fourier}

We start with the Faddeev ansatz (\ref{eq:ansatz_psi_tilde}) for the Fourier transform $\tilde{\psi}_{\rm free}(\kk_1,\ldots,\kk_N)$ of the wavefunction. It may happen that $\tilde{\psi}_{\rm free}$ is identically zero, although the individual contributions $D((\kk_n)_{n\neq i,j})$ are not.  The corresponding scaling exponent is then unphysical and must be disregarded.

For $3+1$ fermions, this happens if the function $D$ is a non identically zero solution of
\be
D(\kk_2,\kk_3)-D(\kk_1,\kk_3)+D(\kk_1,\kk_2)=0 \ \ \ \forall \kk_1,\kk_2,\kk_3
\label{eq:cond3p1}
\ee
From equation (\ref{eq:scaD}), it is expected that $D(\kk_1,\kk_3)$ has a finite limit when $k_3\to +\infty$:
\be
\lim_{k_3\to +\infty} D(\kk_1,\kk_3)=f(\kk_1)
\ee
Taking this limit in equation (\ref{eq:cond3p1}) leads to the correctly antisymmetrised form
\be
D(\kk_1,\kk_2)=f(\kk_1)-f(\kk_2)
\label{eq:D3p1ansatz}
\ee
More generally, differentiating (\ref{eq:cond3p1}) with respect to $\kk_1$ and $\kk_2$, one sees that $D(\kk_1,\kk_2)$ has a vanishing crossed differential, which leads to the same ansatz (\ref{eq:D3p1ansatz}).  The value of the function $f(\kk)$ is actually imposed, up to constant factor, by the rotational symmetry and the scaling invariance.  For a total angular momentum $\ell$ and a scaling exponent $s$, we get
\be
f(\kk)= k^{-(s+\frac{7}{2})} Y_\ell^{\mm}(\hat{\kk})
\ee
where $Y_\ell^{\mm}$ is a spherical harmonic and $\hat{\kk}=\kk/k$ is the direction of $\kk$.  Clearly $f(\kk)$, $D(\kk_1,\kk_2)$ and the final wavefunction have the natural parity $(-1)^\ell$.  Furthermore, as we have seen, the Faddeev component $\FF$ must be a homogeneous polynomial of degree $d$. From the usual power-counting argument in the Fourier transform, we find that $s=d+7/2$, in agreement with equation (\ref{eq:ds}) specialised to $N=4$.  Finally, we take as a particular case $\xx_3=\rr=0$  and we isolate in equation (\ref{eq:compFadFour}) the contribution of the piece $f(\kk_2)$ in $D(\kk_2,\kk_3)$.  We then perform the change of variables $\kk_3=k_2 \kk_3'$ and $\qq=k_2 \qq'$ and formally integrate over $\kk_3$ and $\qq$ the inverse of the energy denominator, which simply pulls out a factor $k_2^4$. We are left with an integral of the form 
\be
\int {\dd}^3k_2 Y_\ell^{\mm}(\hat{\kk}) k_2^{\frac{1}{2}-s} e^{\ii\kk_2\cdot \xx_2}
\label{eq:de_la_forme}
\ee
This must be a homogeneous polynomial in $\xx_2$ of angular momentum $\ell$, of the form (\ref{eq:mono}) with $n$ any natural integer.  Again using a power-counting argument or the change of variable $\kk_2=x_2\kk_2'$, we arrive at the unphysical value of the scaling exponent $s=2n+\ell +\frac{7}{2}$, corresponding to the form (\ref{eq:s4corps}) with $\sigma=1$ and a degeneracy $2\ell+1$.

There is however a little subtlety. In the particular case $(n,\ell)=(0,0)$, that is for a total degree $d=0$ and $s=7/2$, there cannot exist a nonzero fermionic polynomial $P_0(\xx_2,\xx_3)$ of degree zero; the expression (\ref{eq:de_la_forme}) is a constant, as the change of variable $\kk_2=x_2\kk_2'$ shows, and so are the contributions to $\FF(0;\xx_2,\xx_3)$ of the pieces $f(\kk_2)$ and $f(\kk_3)$ of $D(\kk_2,\kk_3)$, which thus exactly cancel.  This was already taken into account in the reasoning above equation (\ref{eq:sgen}) and there is no unphysical solution to disregard.

As a consequence, we obtain a correction to the degeneracy of the scaling exponents of the $3+1$ interaction-sensitive states, 
\be
\bar{\degen}_{\ell,n}^{(\sigma)}=(2\ell+1)\frac{1+\sigma}{2} (1-\delta_{n,0}\delta_{\ell,0})
\label{eq:dbar3p1}
\ee
to be subtracted from the degeneracy $\degen_{\ell,n}^{(\sigma)}$ in equation (\ref{eq:d3p1}).

For $2+2$ fermions, $\tilde{\psi}_{\rm free}(\kk_1,\ldots,\kk_N)$ is identically zero if
\be
D(\kk_2,\kk_4)-D(\kk_2,\kk_3)-D(\kk_1,\kk_4)+D(\kk_1,\kk_3)=0 \ \ \forall \kk_1,\kk_2,\kk_3,\kk_4
\label{eq:cond2p2}
\ee
It is now expected that
\be
\lim_{k_3\to+\infty} D(\kk_1,\kk_3)=f(\kk_1)\ \ \mbox{and}\ \ \lim_{k_1\to+\infty} D(\kk_1,\kk_3)=g(\kk_3)
\ee
As $D$ is not subjected to any exchange symmetry, the functions $f(\kk)$ and $g(\kk)$ are in general independent, but they both tend to zero at large $k$.  Taking the limit $k_1\to +\infty$ and $k_3\to +\infty$ in equation (\ref{eq:cond2p2}), we obtain the ansatz
\be
D(\kk_2,\kk_4)=f(\kk_2)+g(\kk_4)
\label{eq:D2p2ansatz}
\ee
The more direct argument of cross-differentiation of equation (\ref{eq:cond2p2}) with respect to $\kk_2$ and $\kk_4$, which kills all terms but the first one, also leads to the ansatz (\ref{eq:D2p2ansatz}).  The previous $3+1$ reasoning is readily adapted to this case. Due to the rotational symmetry and the scale invariance,
\be
f(\kk)=\alpha k^{-(s+\frac{7}{2})} Y_\ell^{\mm}(\hat{\kk}) \ \ \mbox{and}\ \ g(\kk)=\beta k^{-(s+\frac{7}{2})} Y_\ell^{\mm}(\hat{\kk})
\ee
where $\alpha$ and $\beta$ are arbitrary constants. As the Faddeev components $\FF(0;\xx_2,\mathbf{0})$ and $\FF(0;\mathbf{0},\xx_4)$ must be of the form (\ref{eq:mono}), with $n$ any natural integer, we conclude that the unphysical scaling exponents are of the form (\ref{eq:s4corps}) with a a parity $\sigma=1$ relative to the natural parity, and a degeneracy $2(2\ell+1)$, the extra factor two reflecting the linear independence of $\alpha$ and $\beta$. 

There is here again a little subtlety. In the particular case $(n,\ell)=(0,0)$, the total degree is $d=0$ and the contributions to $\FF(0;\xx_2,\xx_4)$ of the pieces $f(\kk_2)$ and $g(\kk_4)$ in $D(\kk_2,\kk_4)$ are constants proportional to $\alpha$ and $\beta$, so they are not linearly independent.  No extra factor two is required.

As a consequence, we obtain a correction to the degeneracy of the scaling exponents of the $2+2$ interaction-sensitive states, 
\be
\bar{\degen}_{\ell,n}^{(\sigma)}=(2-\delta_{n,0}\delta_{\ell,0})(2\ell+1)\frac{1+\sigma}{2}
\label{eq:dbar2p2}
\ee
to be subtracted from the degeneracy $\degen_{\ell,n}^{(\sigma)}$ in equation (\ref{eq:d2p2}).

The predictions (\ref{eq:d3p1}), (\ref{eq:d2p2}), (\ref{eq:dbar3p1}), (\ref{eq:dbar2p2}) can be tested against the results of reference \cite{Blumeb4}, where the scaling exponents of the interaction-sensitive states of four trapped spin $1/2$ non-interacting fermions were calculated numerically, exhaustively up to some cut-off $s\leq 19/2$ \footnote{For the $(\ell,n,\sigma)=(0,3,+)$ channel of the $2+2$ system, there is a typo in table I of the supplemental material of reference \cite{Blumeb4}, as kindly communicated to us by D\"orte Blume: the scaling exponent of the ideal gas level labeled ``st.\ no.\ 16" should be $\frac{19}{2}$ instead of $\frac{23}{2}$. This is corrected here.}.  As the table \ref{table:vsBlume} shows, there is agreement for $3+1$ fermions and for the unnatural parity states of $2+2$ fermions, but there is disagreement for the natural parity states of $2+2$ fermions. This means that some unphysical states are missed by the above Fourier space reasoning.  This is confirmed in section \ref{subsec:espace_reel}, where it is also exemplified that, surprisingly, the obviously sufficient conditions (\ref{eq:cond3p1}) and (\ref{eq:cond2p2}) to have an unphysical solution are not always necessary.

Still we can give to the unphysical state degeneracies (\ref{eq:dbar3p1}) and (\ref{eq:dbar2p2}) a simple physical picture.  Everything happens as if the particles absent from the set $I$, that is the spin $\uparrow$ particle $i=1$ and the spin $\downarrow$ particle $j=N_\uparrow+1$, were in fact still there and both prepared in the mode $(n,\ell,\mm)=(0,0,0)$. This adds an extra constraint to the modes $(n_i,\ell_i,\mm_i)_{i\in I}$ that can be populated by fermions, a constraint not included in the reasoning above equation (\ref{eq:sgen}). This immediately leads to the occurrence of three types of unphysical solutions:
\begin{itemize}
\item unphysical solutions of type $\uparrow$: one puts one of the spin $\uparrow$ fermions of the set $I$, $2\leq i\leq N_\uparrow$, in the mode
$(0,0,0)$. All the spin $\downarrow$ fermions of the set $I$, $N_\uparrow+2\leq i\leq N$, are put in modes $(n_i,\ell_i,\mm_i)\neq (0,0,0)$.
\item unphysical solutions of type $\downarrow$: one puts one of the spin $\downarrow$ fermions of the set $I$,  $N_\uparrow+2\leq i\leq N$,
in the mode $(0,0,0)$. All the spin $\uparrow$ of the set $I$, $2\leq i\leq N_\uparrow$, are put in modes $(n_i,\ell_i,\mm_i)\neq (0,0,0)$.
\item unphysical solutions of type $\uparrow\downarrow$: one puts one of the spin $\uparrow$ fermions {\sl and} one of the spin $\downarrow$
fermions of the set $I$ in the mode $(0,0,0)$.
\end{itemize}
A natural expectation, that we shall not try to prove here, is that this physical picture applies to all $N$.

\begin{table}[htb]
\begin{center}
\begin{tabular}{|c|c||c|c|c||c|c|c|}
\hline 
$(\ell,n,\sigma)$  & $s_{\ell,n}^{(\sigma)}$ & $\frac{\degen_{\ell,n}^{(\sigma)}}{2\ell+1}$ & $\frac{\bar{\degen}_{\ell,n}^{(\sigma)}}{2\ell+1}$ & $\frac{\degen_{\rm Blume}}{2\ell+1}$ 
& $\frac{\degen_{\ell,n}^{(\sigma)}}{2\ell+1}$ & $\frac{\bar{\degen}_{\ell,n}^{(\sigma)}}{2\ell+1}$ & $\frac{\degen_{\mathrm{Ref. \scriptsize\cite{Blumeb4}}}}{2\ell+1}$ \\
\hline
$(0,0,+)$  & $\frac{7}{2}$  &  0 & 0 & 0 &  1 & 1 & 0 \\
$(0,1,+)$  & $\frac{11}{2}$ &  1 & 1 & 0 &  3 & 2 & 1 \\
$(0,2,+)$  & $\frac{15}{2}$ &  2 & 1 & 1 &  6 & 2$\to$3 & 3 \\
$(0,3,+)$  & $\frac{19}{2}$ &  4 & 1 & 3 & 10 & 2$\to$4 & 6 \\
\hline 
$(1,0,+)$  & $\frac{9}{2}$  &  1 & 1 & 0 &  2 & 2 & 0 \\
$(1,1,+)$  & $\frac{13}{2}$ &  3 & 1 & 2 &  6 & 2$\to$3 & 3 \\
$(1,2,+)$  & $\frac{17}{2}$ &  6 & 1 & 5 &  12 & 2$\to$4 & 8 \\
\hline
$(1,0,-)$  & $\frac{11}{2}$  & 1 & 0 & 1 &  1 & 0 & 1 \\
$(1,1,-)$  & $\frac{15}{2}$ &  2 & 0 & 2 &  3 & 0 & 3 \\
$(1,2,-)$  & $\frac{19}{2}$ &  4 & 0 & 4 &  6 & 0 & 6\\
\hline
$(2,0,+)$  & $\frac{11}{2}$ &  1 & 1 & 0 &  3 & 2 & 1 \\
$(2,1,+)$  & $\frac{15}{2}$ &  4 & 1 & 3 &  9 & 2$\to$3 & 6 \\
$(2,2,+)$  & $\frac{19}{2}$ &  8 & 1 & 7 & 18 & 2$\to$3 & 15 \\
\hline
$(2,0,-)$  & $\frac{13}{2}$ &  1 & 0 & 1 &  2 & 0 & 2 \\
$(2,1,-)$  & $\frac{17}{2}$ &  3 & 0 & 3 &  6 & 0 & 6 \\
\hline
\end{tabular}
\end{center}
\caption{For $3+1$ fermions (left) and $2+2$ fermions (right), values and degeneracies of the scaling exponents of the interaction-sensitive states up to $s=19/2$. The column $\degen_{\ell,n}^{(\sigma)}$ corresponds to the bare degeneracies (\ref{eq:d3p1}) and (\ref{eq:d2p2}). When subtractively corrected by the degeneracies of the unphysical solutions given in the column $\bar{\degen}_{\ell,n}^{(\sigma)}$, it agrees with the numerical results of reference \cite{Blumeb4} reported in the column $\degen_{\mathrm{Ref. \scriptsize\cite{Blumeb4}}}$ (see our footnote $\parallel$).  The values of $\bar{\degen}_{\ell,n}^{(\sigma)}$ are given by the Fourier space predictions (\ref{eq:dbar3p1}) and (\ref{eq:dbar2p2}), corrected if necessary (and as indicated by an arrow) by the real space predictions of section \ref{subsec:espace_reel}.  The parity is $\sigma(-1)^\ell$, $\sigma=\pm$ being relative to the natural parity $(-1)^\ell$.}
\label{table:vsBlume}
\end{table}

\subsection{An investigation in real space for $N=4$}
\label{subsec:espace_reel}

The previous reasoning in Fourier space, though elegant, is formal. It involves integrals with arbitrarily severe infrared divergences, see for example (\ref{eq:de_la_forme}), since $s$ can be arbitrarily large and positive. To believe in this reasoning, it is essential to extract a well defined prediction for the generating polynomial $P_0((\xx_k)_{k\in  I})$ of the unphysical solutions, and to check explicitly, by manipulating polynomials in real space, that the corresponding Faddeev ansatz vanishes.

We shall use two main recipes to obtain finite generating polynomials $P_0$ from the diverging Fourier space integrals. First, we can pull out infinite constants, since $P_0$ is defined up to a global factor. Second, we can use analytic continuation.  Here, we exemplify the procedure for $3+1$ fermions in the manifold $\ell=1$, $n=2$ and $\sigma=+1$. According to the Fourier space reasoning, there should be a single unphysical solution of azimuthal quantum number $\mm=0$. The corresponding polynomial $P_0(\xx_2,\xx_3)=\FF(0;\xx_2,\xx_3)$, of degree $d=2n+\ell=5$, is given by
\be
P_0(\xx_2,\xx_3)= [-\ii\nabla_{\xx_2} \phi(\xx_2,\xx_3) - (\xx_2 \leftrightarrow \xx_3)]\cdot \eee_z
\label{eq:P0nabla}
\ee
where $\eee_z$ is the unit vector along the quantization axis $z$ and where the function $\phi$ is
\be
\phi(\xx_2,\xx_3)=\int \frac{{\dd}^3q {\dd}^3k_2 {\dd}^3k_3}{(2\pi)^{12}}  k_2^{-(d+8)} \frac{e^{\ii(\kk_2\cdot\xx_2+\kk_3\cdot\xx_3)}}{\frac{\hbar^2}{2m_\uparrow}[k_2^2+k_3^2+t(\kk_2+\kk_3)^2+
\frac{q^2}{1-t}]}
\label{eq:phi}
\ee
as it results from equation (\ref{eq:compFadFour}) and a differentiation with respect to $\xx_2$ under the integral sign.  First, we transform (\ref{eq:phi}) only using scaling laws and scale invariances. At fixed $\kk_2$, we perform the change of variable $\kk_3=\kk_3'-\frac{t}{1+t}\kk_2$ to make the energy denominator rotationally invariant. Second we set $\kk_2=(1+t)\kk_2'$ and $\kk_3'=(1+2t)^{1/2}\kk_3''$ so that $k_2'$ and $k_3''$ have identical coefficients in the energy denominator. This leads to the introduction of modified coordinates:
\be
\XX=(1+t)\xx_2-t \xx_3 \ \ \mbox{and}\ \ \ \YY=(1+2t)^{1/2}\xx_3
\ee
We integrate over $\qq$ using a scaling law, $\int \frac{{\dd}^3q}{q^2+Q^2}\propto Q$ for $Q>0$, as the change of variable $\qq=Q\qq'$ shows; this amounts to extracting a diverging constant factor. Integrating over the directions of $\kk_2'$ and $\kk_3''$ and dropping the primes for simplicity, we are left with
\be
\phi(\xx_2,\xx_3)\propto \!\!\int_0^{+\infty} \!\!\! {\dd}k_2 {\dd}k_3 \frac{k_3(k_2^2+k_3^2)^{1/2}}{XY k_2^{d+7}} [\cos(k_2X-k_3Y)-\cos(k_2X+k_3Y)]
\ee
We move to polar coordinates, $(k_3,k_2)=(\rho \cos\theta,\rho\sin\theta)$ to again take advantage of scale invariance: in the integral over $\rho$ involving the first/second cosine term, we perform the change of variable $\rho=\rho'/|X\sin\theta\mp Y\cos\theta|$ and we pull out a common infinite constant factor $\int_{\mathbb{R}^+} \frac{{\dd}\rho'}{\rho'^{d+4}}\cos\rho'$. As $d+3$ is even, we can remove the absolute values and we are left with
\be
\phi(\xx_2,\xx_3)\propto\!\! \int_0^{\pi/2}\!\!\! {\dd}\theta 
\frac{\cos\theta[(X\sin\theta\!-\!Y\cos\theta)^{d+3}\!-\!(X\sin\theta\!+\!Y\cos\theta)^{d+3}]}{XY\sin^{d+7}\theta}
\label{eq:phifin}
\ee
When expanded, the expression in between square brackets only involves odd powers of $\sin\theta$ and of $\cos\theta$. Eliminating the cosine with $\cos^2\theta=1-\sin^2\theta$, we are left with integrals over $\theta$ of the form $f(-n)$, $n\in \mathbb{N}^*$ and
\be
f(z)\equiv \int_0^{\pi/2} {\dd}\theta \sin^{2z+1}\theta
\ee
For $\Re z>-1$ this integral is convergent and given by  $\displaystyle f(z)=\frac{\pi^{1/2}}{2} \frac{\Gamma(z+1)}{\Gamma(z+\frac{3}{2})}$. By the usual analytic continuation of Euler's Gamma function, one can extend $f(z)$ to $\mathbb{C}\setminus \mathbb{R}$, where it can also be written as
\be
f(z) = \frac{\pi^{1/2}}{2\tan(\pi z)} \frac{\Gamma(-z-\frac{1}{2})}{\Gamma(-z)}
\label{eq:fet}
\ee
thanks to Euler's reflection formula $\Gamma(z)\Gamma(1-z)=\pi/\sin(\pi z)$. Unfortunately, this still has poles at the negative integers.  As we are allowed to pull out from equation (\ref{eq:phifin}) a constant diverging factor, we divide it by $f(-1)$, and then we regularise the integrals by introducing a tiny imaginary part in the argument of $f$. We now face
\be
A_n \equiv \lim_{\epsilon\to 0} \frac{f(-n+\ii\epsilon)}{f(-1+\ii\epsilon)}
\ee
As the tangent function is periodic of period $\pi$, the troublesome first denominator in equation (\ref{eq:fet}) is canceled out, the poles disappear and we obtain the
recipe
\be
\frac{\int_0^{\pi/2} {\dd}\theta \sin^{-2n+1}\theta}{\int_0^{\pi/2} {\dd}\theta \sin^{-1}\theta} = A_n = \frac{\Gamma(n-\frac{1}{2})}{\pi^{1/2}\Gamma(n)} \ \ \ \forall n\in \mathbb{N}^*
\ee
For $d=5$ this leads to the finite prediction
\bea
\phi(\xx_2,\xx_3)\propto [A_6 Y^6\!+\!A_5 Y^4(7X^2\!-\!4Y^2)\!+\!A_4 Y^2 (7X^4\!-\!21X^2Y^2\!+\!6Y^4) \nonumber \\ \!+\!A_3(X^6\!-\!14X^4Y^2\!+\!21X^2Y^4\!-\!4Y^6)\!+\!A_2(Y^2\!-\!X^2)(X^4\!-\!6X^2Y^2\!+\!Y^4)]
\eea
Turning to the original variables $\xx_2$ and $\xx_3$ and calculating the gradient in equation (\ref{eq:P0nabla}), we obtain an explicit expression for $P_0(\xx_2,\xx_3)$, and then from (\ref{eq:fadfin}) an explicit expression for $\FF(r;\xx_2,\xx_3)$. We can then evaluate the four-body wavefunction when particles $1$ and $4$ are at the same location, say at the origin of coordinates, from (\ref{eq:ansatz_de_Faddeev}):
\bea
\psi_{\rm free}(\mathbf{0},\xx_2,\xx_3,\mathbf{0}) &=& \FF(0;\xx_2,\xx_3)-\FF(x_2;-t\xx_2,\xx_3-t\xx_2) \\
&& +\FF(x_3;-t\xx_3,\xx_2-t\xx_3)
\eea
After lengthy calculations, we find that it is zero at all $\xx_2$ and $\xx_3$. While we have taken $\rr_1=\rr_4=\mathbf{0}$ for simplicity in the above argument, we can also show, after lengthy calculations, that $\psi_{\rm free}(\rr_1,\rr_2,\rr_3,\rr_4)$ is identically zero for all $\rr_i, 1\leq i\leq 4$.  Thus, our $P_0$ obtained from the Fourier space reasoning indeed generates an unphysical solution.

Is this solution the only one, or is there some unphysical solution missed in section \ref{subsec:Fourier}? To answer this question, still in the manifold $\ell=1$, $n=2$ and $\sigma=+1$ for $3+1$ fermions, we write $P_0$ in the most general form
\be
P_0(\xx_2,\xx_3)= \xx_3\cdot\eee_z \left[\sum_{k=0}^5 c_k p_k(\xx_2,\xx_3)\right] - (\xx_2\leftrightarrow\xx_3)
\label{eq:P0dev}
\ee
where the $p_k(\xx,\yy)$ is a basis of rotationally invariant homogeneous polynomials of degree 4, for example $p_0(\xx,\yy)=y^4$, $p_1(\xx,\yy)=x^4$, $p_2(\xx,\yy)=x^2y^2$, $p_3(\xx,\yy)=y^2 (\mathbf{x}\cdot\mathbf{y})$, $p_4(\xx,\yy)=x^2 (\mathbf{x}\cdot\mathbf{y})$, $p_5(\xx,\yy)=(\mathbf{x}\cdot\mathbf{y})^2 $, and the coefficients $c_k$ are unknown. Then, we calculate the Faddeev component and we expand the resulting polynomial $ \psi_{\rm free}(\mathbf{0},\xx_2,\xx_3,\mathbf{0})$ in the same basis, as in equation (\ref{eq:P0dev}), with coefficients $(c_k')_{0\leq k\leq 5}$ linearly related to the $(c_k)_{0\leq k\leq 5}$ {\sl via} a six-by-six matrix $A$ (too long to be given here). Then $\psi_{\rm free}(\mathbf{0},\xx_2,\xx_3,\mathbf{0})$ is identically zero if and only if all the $c_k'$ are zero, that is
\be
A \vec{c}=\vec{0},
\ee
where the vector $\vec{c}$ collects the six unknowns $(c_k)_{0\leq k\leq 5}$. For a mass ratio $0<t<1$ we find that the null space of $A$ is indeed of dimension one\footnote{Interestingly, the null space of $A$ is of dimension 2 for the infinite mass impurity $t=0$ and of dimension 4 for the zero mass impurity $t=1$, leading to spurious unphysical solutions.}, and is spanned by the Fourier space prediction discussed above.

We have systematically searched for unphysical solutions missed by the Fourier space reasoning for $2+2$ fermions in natural parity states, for all the values of $(\ell,n)$ appearing in the table \ref{table:vsBlume}. The strong motivation to do so is to recover the degeneracies obtained numerically in reference \cite{Blumeb4}, which by construction are exempt from unphysical solutions. We use the previous procedure, expanding $P_0(\xx,\yy)$ over a basis of the homogeneous polynomials $p_k(\xx,\yy)$ of angular momentum $\ell$ and degree $2n+\ell$. We restrict to a zero angular momentum along $\eee_z$, multiplying the obtained degeneracy by $2\ell+1$. As we have seen, for $\ell=0$, we take as a basis the set of monomials $x^{2n_1} y^{2n_2} (\xx\cdot\yy)^{n_3}$, with $n_1+n_2+n_3=n$. For $\ell=1$, we take the set $(\xx\cdot\eee_z) x^{2n_1} y^{2n_2} (\xx\cdot\yy)^{n_3}$ and $(\yy\cdot\eee_z) x^{2n_1} y^{2n_2} (\xx\cdot\yy)^{n_3}$, with $n_1+n_2+n_3=n$. For $\ell=2$, we take $[3(\xx\cdot\eee_z)^2-x^2]x^{2n_1} y^{2n_2} (\xx\cdot\yy)^{n_3}$, $[3(\yy\cdot\eee_z)^2-y^2]x^{2n_1} y^{2n_2} (\xx\cdot\yy)^{n_3}$ and $[3(\xx\cdot\eee_z)(\yy\cdot\eee_z)-\xx\cdot\yy]x^{2n_1} y^{2n_2} (\xx\cdot\yy)^{n_3}$, with $n_1+n_2+n_3=n$. From the generating polynomial with arbitrary coefficients $c_k$ in the basis, we calculate the polynomial $\psi_{\rm free}(\mathbf{0},\xx,\mathbf{0}, \yy)$ and expand it with coefficients $c_k'$ in the same basis. This gives the coefficients of the matrix $A$ relating the $c_k'$ to the $c_k$: $\vec{c}\,'=A\vec{c}$. The number of unphysical solutions is equal to the dimension of the null space of $A$.  As indicated by an arrow in the second $\bar{\degen}$ column of the table, this corrects the Fourier space prediction in six cases.  We then obtain agreement with the numerical results of reference \cite{Blumeb4}.

In all cases, we have found that the unphysical solutions predicted by the Fourier space reasoning (amenable to an explicit prediction for $P_0(\xx,\yy)$ by analytical continuation as explained in this section) are in the null space of the matrix $A$. As we now show on a simple example, some elements of the null space are missed due to the fact that the conditions (\ref{eq:cond3p1}) and (\ref{eq:cond2p2}) are sufficient but not necessary.  Let us consider the case of $2+1$ fermions and take $D(\kk)=k^{-(s+2)}=k^{-4}$, which corresponds to the already known $s=2$ unphysical solution.  This does not satisfy the condition equivalent to (\ref{eq:cond3p1}) for $2+1$ fermions,
that is $D(\kk_2)-D(\kk_1)=0$. Still the generating polynomial $P_0$ is a constant, as a power-counting shows:
\be
P_0(\xx)\propto \int {\dd}^3q {\dd}^3k_2 \frac{D(\kk_2)e^{\ii\kk_2\cdot\xx}}{q^2+(1-t^2) k_2^2} \propto \int {\dd}^3k_2 \frac{e^{\ii\kk_2\cdot\xx}}{k_2^3}\propto x^0
\ee
So is the Faddeev component and, due to the fermionic antisymmetry, the whole Faddeev ansatz $\psi_{\rm free}$ is zero\footnote{There is a $s\leftrightarrow -s$ duality due to the evenness of Efimov's transcendental function, see section \ref{subsec:poles}.  As $D(\kk)\propto k^{-(s+2)}$, the dual of $D(\kk)=k^{-4}$ is $D(\kk)=1$.  It corresponds to a negative value $s=-2$, it can not be mapped to a polynomial Faddeev component in real space and it is not acceptable here. However, it does solve the sufficient condition $D(\kk_2)-D(\kk_1)=0$ for a zero Faddeev ansatz. For three identical bosons, the unphysical solution in the sector $\ell=1, \sigma=1$ is $s=3$ \cite{Werner3corps}, corresponding to $D(\kk)\propto\frac{\hat{\kk}\cdot\eee_z}{k^5}$, so its dual $D(\kk)\propto \kk\cdot\eee_z$ obeys the sufficient condition $D(\kk_1)+D(\kk_2)+D(\kk_3)=0$ {\sl restricted} to the subspace $\kk_1+\kk_2+\kk_3=\mathbf{0}$; the unphysical solution in the sector  $\ell=0$ is $s=4$ \cite{Werner3corps}, corresponding to $D(\kk)\propto k^{-6}$, so its dual $D(\kk)\propto k^2$ by no means obeys the sufficient condition $D(\kk_1)+D(\kk_2)+D(\kk_3)=0$, but one can argue that $D(\kk_1)+D(\kk_2)+D(\kk_3)\propto k_1^2+k_2^2+k_3^2$ simplifies with the energy denominator in the bosonic equivalent of equation (\ref{eq:ansatz_psi_tilde}), leading to $\psi_{\rm free}(\rr_1,\rr_2,\rr_3)=0$ except on a set of zero measure, if the Fourier transform is taken in the framework of distributions.}.

\section{Implications for the cluster or virial expansion for the unitary gas}
\label{sec:viriel}

The cluster expansion is an expansion of the pressure of a thermal equilibrium system in powers of the fugacity, that is at a low density or a high temperature relative to the quantum degeneracy threshold.  It is a powerful tool, because it applies even for strongly interacting systems. Recently, the cluster coefficients were accessed experimentally in the unitary spin $1/2$ Fermi gas up to fourth order \cite{virielENS,virielMIT}.  To calculate the cluster coefficients with the harmonic regulator technique \cite{Comtet,Drummond,CastinCanada}, one must determine the interaction-sensitive energy levels of the unitary gas and of the ideal gas in a harmonic trap.  This makes the connection with the present study.  In section \ref{subsec:poles} we show that the interaction-sensitive energy levels of the ideal gas are related to poles $v_n$ of a generalised Efimov transcendental function $\Lambda(s)$, while  the interaction-sensitive energy levels of the unitary gas are related to roots $u_n$ of $\Lambda(s)$.  In section \ref{subsec:optim} we obtain optimised writings of the third and fourth cluster coefficients in terms of sums $\sum_n (e^{-\bar{\omega} u_n} - e^{-\bar{\omega} v_n})$, which allows us to extend the applicability of the numerical calculations of the fourth cluster coefficient of the reference  \cite{Blumeb4} to lower values of $\bar{\omega}\equiv \hbar\omega/(k_B T)$.  In section \ref{subsec:vsBlume}, we produce some explicit results, showing that the conjecture of reference \cite{pasEfim4} for the fourth cluster clearly fails in the $2+2$ fermion sector, and we construct on physical grounds a new, more successful conjecture.

\subsection{The ideal and unitary gas scaling exponents versus the poles and roots of Efimov's transcendental function}
\label{subsec:poles}

We consider now a zero energy $E=0^-$ solution of Schr\"odinger's equation for two-component interacting fermions in free space, with $\uparrow\downarrow$ contact interactions described by the Wigner-Bethe-Peierls contact conditions of equation (\ref{eq:WBP}).  The regular part $\mathcal{A}((\xx_k)_{k\in I})=A_{1 N_\uparrow+1}((\rr_k-\RR_{1N_\uparrow+1})_{k\neq 1, N_\uparrow+1})$ then solves an integral equation \cite{Cras4corps,Ludovic}
\be
M[\mathcal{A}]=a^{-1} \mathcal{A}
\label{eq:genA}
\ee
where the linear operator $M$ does not depend on the scattering length $a$ and the set $I$ is given by the equation (\ref{eq:defI}).  In the unitary limit $a^{-1}=0$ and in the ideal gas limit $a^{-1}=\infty$, the gas is scale invariant and the function $\mathcal{A}$ has some scaling exponent $s_\mathcal{A}$, conveniently defined by a shift of $+1$ in the exponent of equation (\ref{eq:defs}):
\be
\mathcal{A}(\lambda(\xx_k)_{k\in I})= \lambda^{s_\mathcal{A}+1-\frac{3N-5}{2}} \mathcal{A}((\xx_k)_{k\in I})\ \ \ \forall \lambda >0
\label{eq:defsA}
\ee
Inserting in the equation (\ref{eq:genA}) the ansatz $\mathcal{A}((\xx_k)_{k\in I})=R_\mathcal{A}^{s_\mathcal{A}+1-\frac{3N-5}{2}} \Phi(\Omega_\mathcal{A})$ 
where $R_\mathcal{A}$ is the hyperradius
and $\Omega_\mathcal{A}$ are hyperangles parametrising the $(\xx_k)_{k\in I}$, one obtains an implicit equation for $s_\mathcal{A}$, 
\be
\Lambda_\ell^{(\sigma)}(s_\mathcal{A})\stackrel{a^{-1}=0}{=}0 \ \ \ \mbox{or}\ \ \ \Lambda_\ell^{(\sigma)}(s_\mathcal{A})\stackrel{a^{-1}=\infty}{=}\infty
\label{eq:sursA}
\ee
where we could restrict to a subspace of fixed angular momentum $\ell\in \mathbb{N}$ and parity $\sigma (-1)^\ell$, $\sigma=\pm 1$, due to the rotational invariance and the parity invariance.  Formally $\Phi(\Omega_\mathcal{A})$ is the eigenvector of some linear $s_\mathcal{A}$-dependent operator $\mathcal{M}_\ell^{(\sigma)}(s_\mathcal{A})$ with a zero or an infinite eigenvalue, the function $\Lambda_\ell^{(\sigma)}(s_\mathcal{A})$ is the determinant of that linear operator,
\be
\Lambda_\ell^{(\sigma)}(s_\mathcal{A})=\mathrm{det}\ \mathcal{M}_\ell^{(\sigma)}(s_\mathcal{A})
\label{eq:defgenLam}
\ee
and is obviously independent of $a$.  We call it Efimov's transcendental function, because it was calculated analytically by Efimov for $N=3$ \cite{Efimov}, see also references \cite{Werner3corps,Gasaneo,Birse,Rittenhouse,Tignone}.  For $N=4$, it was evaluated numerically, for imaginary values of $s_\mathcal{A}$ only \cite{pasEfim4,CMP}. Importantly, it is an even function of $s_{\mathcal{A}}$.  In what follows, we assume that there is no $N$-body Efimov effect, which leads to known constraints on the mass ratio $m_\uparrow/m_\downarrow$ for $N=3$ \cite{Efimov,Tignone,Kartavtsev} and for $N=4$ \cite{pasEfim4,CMP}. As a consequence, all the roots of $\Lambda_\ell^{(\sigma)}(s_\mathcal{A})$ are real. Considering (\ref{eq:sursA}) we call $(u_{\ell,n}^{(\sigma)})_{n\in\mathbb{N}}$ the set of positive roots of $\Lambda_\ell^{(\sigma)}$ and $(v_{\ell,n}^{(\sigma)})_{n\in\mathbb{N}}$ the set of positive poles of $\Lambda_\ell^{(\sigma)}$, counted with a degeneracy $2\ell+1$, the negative roots and poles being their opposites:
\be
\Lambda_{\ell}^{(\sigma)}(u_{\ell,n}^{(\sigma)}>0)=0 \ \ \mbox{and}\ \ \Lambda_{\ell}^{(\sigma)}(v_{\ell,n}^{(\sigma)}>0)=\infty,\ \ \forall n\in \mathbb{N}
\label{eq:uv}
\ee

The last step is to relate the scaling exponent $s_\mathcal{A}$ in equation (\ref{eq:defsA}) of the regular part $\mathcal{A}$ to the scaling exponent $s$ (\ref{eq:defs}) of the wavefunction $\psi_{\rm free}(\rr_1,\ldots,\rr_N)$. For the unitary gas, denoted by a diacritical sign, the term $\frac{1}{a}$ vanishes in the Wigner-Bethe-Peierls contact condition, so $\psi_{\rm free} \sim r^{-1} \mathcal{A}$ and, thanks to the shift of $+1$ of the exponent in the definition (\ref{eq:defsA}) one simply has
\be
\check{s}\stackrel{a^{-1}=0}{=} s_\mathcal{A}
\ee
For the ideal gas, the term $\frac{1}{a}$ diverges in the Wigner-Bethe-Peierls contact condition, so $\psi_{\rm free} \sim a^{-1} \mathcal{A}$ and
\be
s\stackrel{a^{-1}=\infty}{=} s_\mathcal{A}+1
\ee
Combining the equations (\ref{eq:sursA}) and (\ref{eq:uv}), the general considerations from the SO(2,1) symmetry of the Hamiltonian in section \ref{subsec:SO21}, and the fact that only positive values of $s$ are physical in the absence of $N$-body resonances \cite{WernerCastinSO21,Petrov2003,Felixthese,NishidaTan}, we obtain the general expressions for the SO(2,1) ladders of internal energy levels of the unitary gas and of the ideal gas in terms of the positive roots and the positive poles of Efimov's transcendental function, each energy level being counted with a degeneracy $2\ell+1$:
\be
\check{E}_q^{\rm rel}\stackrel{a^{-1}=0}{=}(u_{\ell,n}^{(\sigma)}+1+2q) \hbar\omega \ \ \ \mbox{and}\ \ \ 
E_q^{\rm rel}\stackrel{a^{-1}=\infty}{=} (v_{\ell,n}^{(\sigma)}+2+2q)\hbar\omega  \ \ \forall q\in \mathbb{N}
\label{eq:suv}
\ee
This remarkable property was noticed and used in reference \cite{CastinCanada} for three bosons and in reference \cite{Gao} for three fermions, but it was not physically interpreted. We have presented here a general physical derivation of this fact, independently of the particle number. Note that (\ref{eq:suv}) includes the unphysical solutions as defined in section \ref{sec:non_physique} because it involves an integral equation (\ref{eq:genA}) ultimately relying on the Faddeev ansatz; these unphysical solutions are {\sl common} to the ideal gas and the unitary gas, because the Faddeev ansatz, being zero, satisfies the Wigner-Bethe-Peierls contact conditions for all values of the scattering length $a$ \cite{CastinCanada}. Whereas the $u_{\ell,n}^{(\sigma)}$ can probably not be determined analytically beyond $N=3$, the $v_{\ell,n}^{(\sigma)}$ can be explicitly obtained from our results of section \ref{sec:le_resultat_en_gros}.

\subsection{Optimized writing of the fourth-order cluster coefficients}
\label{subsec:optim}

In the harmonic regulator method \cite{Comtet}, one performs the cluster or virial expansion for the thermal equilibrium harmonically trapped system. The grand potential of the trapped two-component Fermi gas is by definition
\be
\Omega=-k_B T \ln \left(\sum_{N_\uparrow=0}^{+\infty} \sum_{N_\downarrow=0}^{+\infty} Z_{N_\uparrow,N_\downarrow} z_\uparrow^{N_{\uparrow}}z_\downarrow^{N_\downarrow}\right)
\label{eq:dev1}
\ee
where $Z_{N_\uparrow,N_\downarrow}$ is the canonical partition function of $N_\uparrow+N_\downarrow$ fermions at temperature $T$ in isotropic harmonic traps with a common angular frequency $\omega$ for the $\uparrow$ and $\downarrow$ components, and the fugacities $z_\sigma$ of the components are related to their chemical potentials $\mu_\sigma$ by $z_\sigma=e^{\beta \mu_\sigma}$, with $\beta=(k_B T)^{-1}$. In the low density, non-degenerate limit $\mu_\sigma\to -\infty$ at fixed temperature, that is $z_\sigma\to 0$, one performs the so-called cluster expansion of the grand potential \cite{Huang}:
\be
\Omega=-k_B T Z_1 \sum_{N_\uparrow=0}^{+\infty} \sum_{N_\downarrow=0}^{+\infty} B_{N_\uparrow,N_\downarrow} z_\uparrow^{N_{\uparrow}}z_\downarrow^{N_\downarrow}
\label{eq:dev2}
\ee
where $Z_1=Z_{1,0}=Z_{0,1}$, the single fermion partition function in the trap, is given by
\be
Z_1=\frac{1}{[2\sinh(\bar{\omega}/2)]^{3}} \ \ \ \mbox{with}\ \ \ \bar{\omega}\equiv \beta \hbar \omega
\ee
It is convenient to restrict to the differences $\Delta Z_{N_\uparrow,N_\downarrow}$ and $\Delta B_{N_\uparrow,N_\downarrow}$ between the interacting gas and the ideal gas values of $Z_{N_\uparrow,N_\downarrow}$ and $B_{N_\uparrow,N_\downarrow}$: the ideal gas values are elementary to calculate, and in the differences, the contribution of the interaction-insensitive states exactly cancel.  Also, one can use the separability of the centre of mass and the relative motion of the particles to restrict to the partition functions of the internal energy levels,
\be
\Delta Z_{N_\uparrow,N_\downarrow}= Z_1  \Delta Z^{\rm rel}_{N_\uparrow,N_\downarrow}
\ee
These internal or relative energies are what is ultimately calculated, see the equations (\ref{eq:Eq}), (\ref{eq:jolie_forme}) and (\ref{eq:suv}) and the references \cite{Drummond,Blumeb4}.  Expanding the equation (\ref{eq:dev1}) in powers of the fugacities and equating the coefficients of $z_\uparrow^{N_\uparrow}z_\downarrow^{N_\downarrow}$ to those of the equation (\ref{eq:dev2}), one obtains up to fourth order:
\bea
\label{eq:DB11}
\Delta B_{1,1} &=& \Delta Z_{1,1}^{\rm rel} \\
\label{eq:DB21}
\Delta B_{2,1} &=& \Delta Z_{2,1}^{\rm rel} -Z_1 \Delta B_{1,1}\\
\label{eq:DB31}
\Delta B_{3,1} &=& \Delta Z_{3,1}^{\rm rel} - Z_1 Z_{2,0}^{\rm rel} \Delta B_{1,1} - Z_1 \Delta B_{2,1} \\
\label{eq:DB22}
\Delta B_{2,2} &=& \Delta Z_{2,2}^{\rm rel} - Z_1^2 \Delta B_{1,1} - Z_1 \left(\frac{1}{2}  \Delta B_{1,1}^2 + \Delta B_{2,1} + \Delta B_{1,2}\right)
\eea
At any given order, we have recursively used the relations obtained at a lower order to eliminate partition functions $\Delta Z^{\rm rel}$ in terms of cluster
coefficients $\Delta B$.  The cluster coefficients with $N_\uparrow < N_\downarrow$ are obtained by exchanging the roles of $\uparrow$ and $\downarrow$ in the above expressions.  Note that $\Delta B_{N_\uparrow,0}=\Delta B_{0,N_\downarrow}=0$. Also the ideal gas values $B_{N_\uparrow,N_\downarrow}^{a=0}$ are zero except if $N_\uparrow=0$ or $N_\downarrow=0$. Last, from a use of the centre of mass and relative quantum numbers of two $\uparrow$ fermions as explained around equation (\ref{eq:s31relcom}), one has
\be
Z_{2,0}^{\rm rel} =\sum_{\ell_{\rm rel}\in 2\mathbb{N}+1} \sum_{n_{\rm rel}=0}^{+\infty}  (2\ell_{\rm rel}+1) 
e^{-(2n_{\rm rel}+\ell_{\rm rel}+3/2)\bar{\omega}}=
\frac{e^{-3\bar{\omega}/2}(1+3e^{2\bar{\omega}})}{(2\sinh \bar{\omega})^3}
\ee

From now on, the interacting two-component Fermi gas is taken in the unitary limit $a^{-1}=0$.  For $1+1$ fermions the scaling exponents in equations (\ref{eq:defs}) and (\ref{eq:Eq}) are respectively $\check{s}=-1/2$ and $s=1/2$ in the unitary and non-interacting limits so, from equation (\ref{eq:DB11}),
\be
\Delta B_{1,1}=\sum_{q\in\mathbb{N}} \left[e^{-(2q+1/2)\bar{\omega}} - e^{-(2q+3/2)\bar{\omega}}\right] = \frac{1}{2\cosh (\bar{\omega}/2)}
\label{eq:DB11expli}
\ee
For higher order cluster coefficients, the goal is to obtain {\sl optimized} writings in terms of the following sums,
\be
\label{eq:defS}
S_{N_\uparrow,N_\downarrow} \equiv \sum_{n,\ell,\sigma} (2\ell+1) \left[e^{-u_{\ell,n}^{(\sigma)}\bar{\omega}}-e^{-v_{\ell,n}^{(\sigma)}\bar{\omega}}\right]
\ee
where the roots $u_{\ell,n}^{(\sigma)}$ and poles $v_{\ell,n}^{(\sigma)}$ of Efimov's transcendental function (\ref{eq:defgenLam}) for $N_\uparrow+N_\downarrow$ fermions are defined in section \ref{subsec:poles}, and since this sum only involves interaction-sensitive states, the relative parity $\sigma$ is $+1$ for $N=3$ and $\pm 1$ for $N=4$.  The motivation is that these sums evoke the sums of residues resulting from the application of Cauchy's theorem to a contour integration of the functions $s_\mathcal{A}\mapsto e^{-s_\mathcal{A}\bar{\omega}} \frac{\rm d}{{\rm d}s_\mathcal{A}} \ln\Lambda_{\ell}^{(\sigma)}(s_\mathcal{A})$ \cite{CastinCanada}, which suggests that they can be expressed in terms of an integral of these functions on the imaginary axis where they can be calculated even for $N=4$ \cite{pasEfim4,CMP}.  Importantly, these sums $S_{N_\uparrow,N_\downarrow}$ include the unphysical solutions discussed in section \ref{sec:non_physique}.  This is in marked contrast with $\Delta Z_{2,1}^{\rm rel}$, $\Delta Z_{3,1}^{\rm rel}$ and $\Delta Z_{2,2}^{\rm rel}$  where one can indifferently include or exclude the unphysical solutions for $N=3$ and $N=4$, since they are common to the ideal gas and the unitary gas and cancel out in $\Delta Z_{2,1}^{\rm rel}$, $\Delta Z_{3,1}^{\rm rel}$ and $\Delta Z_{2,2}^{\rm rel}$.

To express the $\Delta B$ in terms of the sums $S$, we start from
\be
\Delta Z_{N_\uparrow,N_\downarrow}^{\rm rel} = \sum_{n,\ell,q,\sigma} (2\ell+1) \left[e^{-(u_{\ell,n}^{(\sigma)}+1+2q)\bar{\omega}}-
e^{-(v_{\ell,n}^{(\sigma)}+2+2q)\bar{\omega}}\right]
\ee
then we use a plus-minus trick, writing $\exp[-(v_{\ell,n}^{(\sigma)}+2+2q)\bar{\omega}]=\exp[-(v_{\ell,n}^{(\sigma)}+1+2q)\bar{\omega}]-(\exp\bar{\omega}-1)\exp[-(v_{\ell,n}^{(\sigma)}+2+2q)\bar{\omega}]$ 
\be
\Delta Z_{N_\uparrow,N_\downarrow}^{\rm rel}=\frac{S_{N_\uparrow,N_\downarrow}}{2\sinh\bar{\omega}} 
+(e^{\bar{\omega}}-1) Z_{N_\uparrow,N_\downarrow}^{\rm{rel,int.sens. with\, unphys. sol.}}
\ee
where $Z_{N_\uparrow,N_\downarrow}^{\rm{rel,int.sens. with\, unphys. sol.}}=\sum_{n,\ell,\sigma,q} (2\ell+1) \exp[-(v_{\ell,n}^{(\sigma)}+2+2q)\bar{\omega}]$ is the partition function of the interaction-sensitive states of the relative motion of trapped $2+1$ fermions, $3+1$ fermions or $2+2$ fermions {\sl including} the unphysical solutions. This partition function is known from  the equation (\ref{eq:jolie_forme}): it is equal to $Z_1^{\ell=0}$ times the partition function of respectively one trapped particle $Z_1$, two trapped $\uparrow\uparrow$ fermions $Z_{2,0}$ or two trapped $\uparrow\downarrow$ non-interacting particles $Z_{1,1}^{a=0}=Z_1^2$. Here $Z_1^{\ell=0}=e^{-\bar{\omega}/2}/[2\sinh \bar{\omega}]$ is the single particle partition function restricted to the $\ell=0$ states and accounts for the term $(2q+\frac{3}{2})\hbar \omega$ in the equation (\ref{eq:jolie_forme}).  Since
\be
(e^{\bar\omega}-1)Z_1^{\ell=0} - \Delta B_{1,1}=0
\ee
this leads to the reduced forms\footnote{In evaluating $\sum_{n,\ell,\sigma,q} (2\ell+1)\exp[-(v_{\ell,n}^{(\sigma)}+2+2q)\bar{\omega}]$ we include unphysical solutions. Thus, the corresponding sum $S_{N_\uparrow,N_\downarrow}$ must also include unphysical solutions. An alternative choice would be to exclude the unphysical solutions in both sums; this would be inconvenient because we do not know all unphysical solutions for $2+2$ fermions.}
\bea
\label{eq:DB21red}
\Delta B_{2,1} &=& \frac{S_{2,1}}{2\sinh \bar{\omega}} \\
\label{eq:DB31red}
\Delta B_{3,1} &=& \frac{S_{3,1}}{2\sinh \bar{\omega}} - Z_1 \Delta B_{2,1}\\
\label{eq:DB22red}
\Delta B_{2,2} &=& \frac{S_{2,2}}{2\sinh \bar{\omega}}- Z_1 \left(\frac{1}{2}  \Delta B_{1,1}^2 + \Delta B_{2,1} + \Delta B_{1,2}\right) 
\eea
To obtain the cluster coefficients of the spatially homogeneous gas, one must calculate the (finite) limit of the $\Delta B$ when $\bar{\omega}\to 0$. These forms are then optimised in the sense that one has got rid in the $\Delta B$ of the term $\propto Z_1$ diverging as $1/\bar{\omega}^3$ for $N=3$, and of the terms $\propto Z_1^2$ diverging as $1/\bar{\omega}^6$ for $N=4$.

\subsection{Application to the numerical data of reference \cite{Blumeb4} and test of old and new conjectures}
\label{subsec:vsBlume}

For $2+1$ unitary fermions, an application of Cauchy's theorem to the reduced form (\ref{eq:DB21red}) as in reference \cite{CastinCanada} gives the exact result:
\be
\label{eq:DB21ex}
\Delta B_{2,1} = \sum_{\ell\in\mathbb{N}} \left(2\ell+1\right)\int_\mathbb{R} \frac{{\dd}S}{2\pi} \frac{\sin(\bar{\omega}S)}{2\sinh\bar{\omega}} \frac{{\dd}}{{\dd}S} 
\left[\ln \Lambda_\ell(\ii S)\right]
\ee
For $3+1$ or $2+2$ unitary fermions, it is not understood yet how the terms $\propto Z_1$ in equations (\ref{eq:DB31red}) and (\ref{eq:DB22red}), which diverge for $\bar{\omega}\to 0^+$ contrarily to $\Delta B_{N_\uparrow,N_\downarrow}$, can be compensated by poles of the logarithmic derivative of the corresponding Efimov's transcendental function\footnote{Alternatively to such unexpected poles, one can invoke a nonzero contribution of the infinite-radius quarters of circle used to connect the contour integration enclosing the poles on the real axis to the integral along the imaginary axis. Due to the continuous spectrum of the operators $\mathcal{M}_\ell^{(\sigma)}(\ii S)$, one can also suspect the existence of branch cuts for the function $z\mapsto \frac{\dd}{\dd z}[\ln \Lambda_{\ell}^{(\sigma)}(z)]$ in the complex plane; turning around those branch cuts in the contour integration would then lead to extra contributions. Our current knowledge of the function $\Lambda_{\ell}^{(\sigma)}(z)$, limited to $z\in \ii \mathbb{R}$, does not allow to settle the problem.}.  Simply the following conjectured values were proposed in reference \cite{pasEfim4}, by an abrupt generalisation of equation (\ref{eq:DB21ex}) with no justification:
\be
\Delta B_{N_\uparrow,N_\downarrow}^{\rm old\, conj} = I_{N_\uparrow,N_\downarrow}
\label{eq:DBconj}
\ee 
Here $I_{N_\uparrow,N_\downarrow}$ is the following imaginary axis integral, shown to be finite in reference \cite{pasEfim4}:
\be
I_{N_\uparrow,N_\downarrow}\equiv \sum_{\ell\in\mathbb{N},\sigma=\pm 1} (2\ell+1)\int_\mathbb{R} \frac{{\dd}S}{2\pi} \frac{\sin(\bar{\omega}S)}{2\sinh\bar{\omega}} \frac{{\dd}}{{\dd}S} \left[\ln \Lambda_\ell^{(\sigma)}(\ii S)\right]
\label{eq:IAI}
\ee
where $\Lambda_{\ell}^{(\sigma)}$ is given by the equation (\ref{eq:defgenLam}) and the corresponding linear operators $\mathcal{M}_\ell^{(\sigma)}(\ii S)$ are given in reference \cite{CMP} for $3+1$ fermions and in reference \cite{pasEfim4} for $2+2$ fermions.  The resulting value of the fourth cluster coefficient $\Delta b_4$ of the spatially homogeneous system for $m_\uparrow=m_\downarrow$ is however in complete contradiction with the experimental results \cite{virielENS,virielMIT}, even for its sign.

We make here a more detailed test of the conjecture (\ref{eq:DBconj}). From the scaling exponents $\check{s}_{\ell,n}^{(\sigma)}$ and $s_{\ell,n}^{(\sigma)}$ of the interaction-sensitive states of the unitary and ideal four-particle systems calculated numerically up to the cut-off $\check{s}=19/2$ in reference \cite{Blumeb4}, one can accurately calculate $\Delta B_{3,1}$ and $\Delta B_{2,2}$ for not too small $\bar{\omega}$.  To evaluate $\Delta B_{2,1}$, which appears in the expressions of $\Delta B_{3,1}$ and $\Delta B_{2,2}$, we do not use a numerically calculated three-body spectrum, but rather the exact expression (\ref{eq:DB21ex}). 

\begin{figure}[htb]
\begin{center}
\includegraphics[width=7cm,clip=]{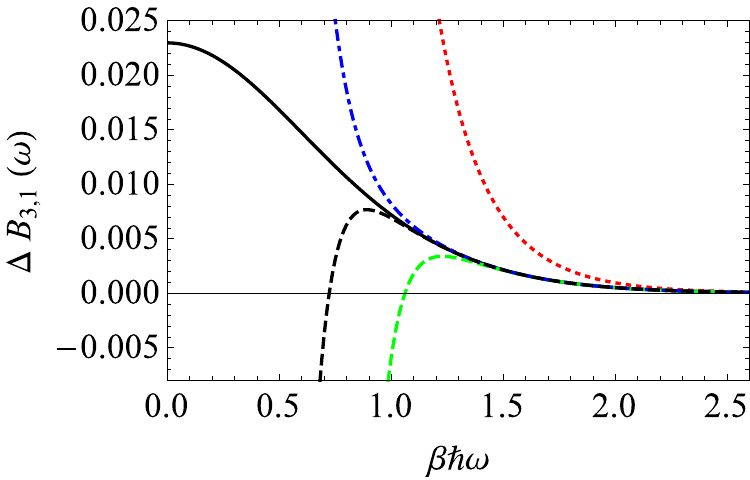}
\includegraphics[width=7cm,clip=]{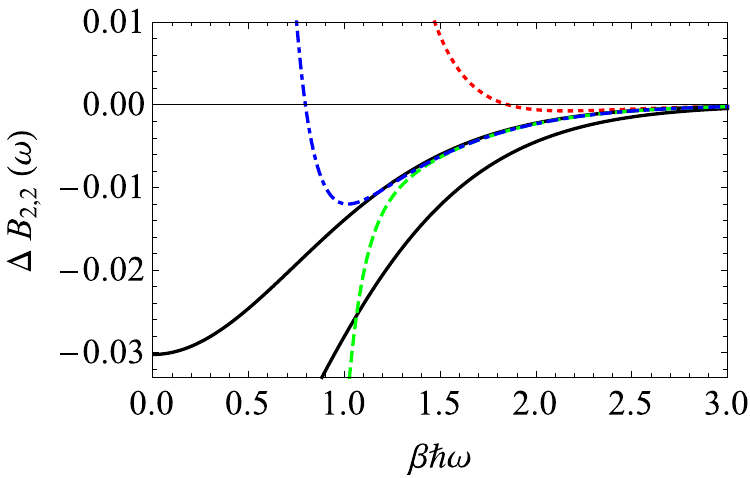}
\end{center}
\caption{Fourth-order cluster coefficients $\Delta B_{3,1}$ (left panel) and $\Delta B_{2,2}$ (right panel) of a harmonically trapped two-component unitary Fermi gas with equal masses $m_\uparrow=m_\downarrow$, as functions of $\beta \hbar\omega$, where $\omega$ is the trapping angular frequency, $\beta=1/(k_B T)$ and $T$ is the temperature.  Green dashed lines: from the original formulas (\ref{eq:DB31}) and (\ref{eq:DB22}) and the numerical four-body spectrum of reference \cite{Blumeb4}.  Red dotted lines: from an incorrect use of the optimised formulas (\ref{eq:DB31red}) and (\ref{eq:DB22red}), only the physical solutions being included.  Blue dashed-dotted lines: from a correct use of  (\ref{eq:DB31red}) and (\ref{eq:DB22red}), both the physical and the unphysical solutions being included and subjected to a cut-off.  Black dashed line (left panel only): idem, except that all unphysical solutions are included, with a degeneracy (\ref{eq:dbar3p1}).  All these lines diverge at low $\bar{\omega}$, because the four-body spectrum in reference \cite{Blumeb4} is calculated up to some cut-off [in our calculations, any three-body cut-off is avoided thanks to the exact expression (\ref{eq:DB21ex})].  Black solid lines: the conjectured values; left panel: the old (\ref{eq:DBconj}) and new (\ref{eq:newconj31}) conjectures coincide and are in good agreement with the numerics; right panel: the old conjecture (\ref{eq:DBconj}) (lower black solid line) disagrees with the numerics, whereas the new conjecture (\ref{eq:newconj22}) (upper black solid line) is in good agreement.}
\label{fig:blvscj}
\end{figure}

Various results for $\Delta B_{3,1}$ and for $\Delta B_{2,2}$, corresponding to various expressions given in the present paper, are plotted as functions of $\bar{\omega}=\beta\hbar\omega$ in figure \ref{fig:blvscj}. The green dashed lines correspond to the original formulas (\ref{eq:DB31}) and (\ref{eq:DB22}).  They start diverging at $\bar{\omega}\lesssim 1.3$, due to the finite cut-off. The red dotted lines correspond to an {\sl incorrect} application of the optimised formulas (\ref{eq:DB31red}) and (\ref{eq:DB22red}), that is including in $S_{3,1}$ and $S_{2,2}$ only the physical solutions obtained by the reference \cite{Blumeb4}; the unphysical solutions are missing, and the red dotted curves start diverging at {\sl larger} values of $\bar{\omega}$, $\bar{\omega}\simeq 2.2$.  The blue dashed-dotted lines correspond to the {\sl correct} use of the optimised formulas (\ref{eq:DB31red}) and (\ref{eq:DB22red}): one includes in $S_{3,1}$ and $S_{2,2}$ both the physical and the unphysical solutions up to the cut-off, where the degeneracies of the unphysical solutions are obtained as the difference of the bare degeneracies $\degen_{\ell,n}^{(\sigma)}$ given by the equations (\ref{eq:d3p1}) and (\ref{eq:d2p2}) with the degeneracies of the physical solutions obtained numerically in reference \cite{Blumeb4}. As expected, the blue dashed-dotted lines start diverging at {\sl smaller} values of $\bar{\omega}$, $\bar{\omega}\simeq 1$. The black dashed line for $\Delta B_{3,1}$ is an improvement over the blue dashed-dotted line: one sums over all unphysical solutions, without cut-off, assuming that their degeneracy $\bar{\degen}_{\ell,n}^{(\sigma)}$ is given by the Fourier space reasoning (\ref{eq:dbar3p1}) of section \ref{subsec:Fourier}; the divergence then starts at an even smaller value of $\bar{\omega}$, $\bar{\omega}\lesssim 0.9$. Such improvement can not be performed for $\Delta B_{2,2}$, because the degeneracy $\bar{\degen}_{\ell,n}^{(\sigma)}$ predicted by the equation (\ref{eq:dbar2p2}) is an underestimate, see the table \ref{table:vsBlume} and the section \ref{subsec:espace_reel}.

The conjectured values (\ref{eq:DBconj}) are plotted as black solid lines (lower black solid line on the right panel).  For $3+1$ fermions, they essentially agree with the blue dashed-dotted line up to the point of its cut-off induced divergence. This leaves the possibility that the conjecture is correct for $\Delta B_{3,1}$.  Incidentally, its $\bar{\omega}\to 0^+$ limit $\Delta B_{3,1}^{\rm old\, conj}(0^+)= 0.02297(4)$ \cite{pasEfim4} is close to the prediction $0.025$ of the approximate diagrammatic method of reference \cite{Levinsen}  (this value was communicated to us privately by Jesper Levinsen).  For $2+2$ fermions, the conjectured values (\ref{eq:DBconj}) clearly disagree with the blue dashed-dotted lines even in the cut-off unaffected region $\bar{\omega} \geq 1$ and with the approximate value $\Delta B_{2,2}(0^+)=-0.036$ of reference \cite{Levinsen}, by a factor close to $2$.  The conjecture (\ref{eq:DBconj}) is thus invalidated for $\Delta B_{2,2}$. 

Let us now construct a less arbitrary conjecture than (\ref{eq:DBconj}) for the fourth-order cluster coefficients, building on our physical understanding.  The imaginary axis integrals (\ref{eq:IAI}) have a finite limit when $\omega\to 0^+$ \cite{pasEfim4}. They must differ from the expected sum of residues $S_{N_\uparrow,N_\downarrow}/(2\sinh\bar{\omega})$, that diverges when $\bar{\omega}\to 0^+$ as we have seen, by counter-terms  $C_{N_\uparrow,N_\downarrow}$ of unelucidated mathematical origin:
\be
\label{eq:CNN}
I_{N_\uparrow,N_\downarrow}= \frac{S_{N_\uparrow,N_\downarrow}}{2\sinh\bar{\omega}}-C_{N_\uparrow,N_\downarrow}
\ee
The new feature of the $3+1$ and $2+2$ fermion problem with respect to the $2+1$ one is that $\mathcal{M}_\ell^{(\sigma)}(\ii S)$ in equation~(\ref{eq:defgenLam}), rather than being a finite size matrix of a purely discrete spectrum, is an operator with a continuous spectrum.  We then expect that this continuous spectrum is at the origin of the sought counter-terms.  Crucially, the continuous spectrum can be interpreted in terms of {\sl decoupled asymptotic objects} (DAOs), emerging for large amplitude oscillations of the four fermions in the trap or equivalently for eigenstates with large quantum numbers \cite{pasEfim4,CMP}.  For such asymptotic states, one indeed expects that the $N_\uparrow+N_\downarrow$ particles split into single particles or groups of strongly correlated particles, that we call DAOs and that do not interact with the other groups or particles because they have high amplitude relative motions.  The relevant DAOs and their spectral properties are presented in the table \ref{table:DAO}. The partition functions $Z_{\rm DAO}$ of the DAOs in the trap, or more precisely the relative partition functions $Z^{\rm rel}_{\rm DAO}=Z_{\rm DAO}/Z_1$ after removal of the centre of mass, are easy to calculate since by essence the DAOs do not interact.  They will be assumed to provide the counter-terms in the new conjecture.

The $3+1$ fermion case is the simplest one. As a whole, the continuous spectra of the operators $\mathcal{M}_\ell^{(\pm)}(\ii S)$ are composed of branches $k\mapsto\Lambda^{\uparrow\uparrow\downarrow}_L(\ii k)$ of degeneracy $2L+1$, $k\in\mathbb{R}$ and $L\in\mathbb{N}$. Each branch corresponds to two DAOs: a cluster of neighbouring, strongly correlated $\uparrow\uparrow\downarrow$ fermions of angular momentum $L$ and a decoupled $\uparrow$ atom of orbital angular momentum compatible with the total spin $\ell$.  We call the $\uparrow\uparrow\downarrow$ cluster a {\sl triplon}; its wavefunction has scaling exponents $s$ given by the positive roots $u_{L,n}^{\uparrow\uparrow\downarrow}$ of $\Lambda^{\uparrow\uparrow\downarrow}_L(s)$; in the trap, it moves as a free particle with an {\sl internal} structure of energy levels $(u_{L,n}^{\uparrow\uparrow\downarrow}+1+2q)\hbar\omega$, $(L,n,q)\in\mathbb{N}^3$.  Its ideal gas counterpart and the difference, indicated as above by the letter $\Delta$, of their relative partition functions $\Delta Z_{\uparrow\uparrow\downarrow\,\mathrm{triplon}}^{\rm rel}$, are given in the table \ref{table:DAO}.  We expect that the counter-term $C_{3,1}$ to be subtracted from $S_{3,1}/(2\sinh\bar{\omega})$ is the difference of the relative atom+triplon partition functions in the unitary and ideal gases, 
\be
C_{3,1}=\Delta Z_{\uparrow\,\mathrm{atom}+\uparrow\uparrow\downarrow\,\mathrm{triplon}}^{\rm rel}
\ee
hence the new conjecture
\be
I_{3,1}=\frac{S_{3,1}^{\rm new\, conj}}{2\sinh\bar{\omega}}
-\Delta Z_{\uparrow\,\mathrm{atom}+\uparrow\uparrow\downarrow\,\mathrm{triplon}}^{\rm rel}
\label{eq:C31}
\ee
\begin{table}[htb]
\begin{center}
\begin{tabular}{|c|c|c|c|c|}
\hline
DAO &  $\epsilon_{\rm int}^{\rm unit}/\hbar\omega$ & $\epsilon_{\rm int}^{\rm ideal}/\hbar\omega$ & $\Delta Z_{\rm DAO}^{\rm rel}$ \\
\hline
$\uparrow$ or $\downarrow$ atom & $0$ & $0$ & $0$ \\
$\uparrow\downarrow$  pairon & $2q+\frac{1}{2}$ & $2q+\frac{3}{2}$ & $\Delta Z_{1\, \rm pairon}^{\rm rel}= \Delta B_{1,1}$ \\
$\uparrow\uparrow\downarrow$  triplon & $u^{\uparrow\uparrow\downarrow}_{L,n}+2q+1$ & $v^{\uparrow\uparrow\downarrow}_{L,n}+2q+2$ &
$\Delta Z_{\uparrow\uparrow\downarrow\, \rm triplon}^{\rm rel}=\Delta Z_{2,1}^{\rm rel}-Z_1\Delta B_{1,1}=\Delta B_{2,1}$ \\
$\uparrow\downarrow\downarrow$  triplon & $u^{\uparrow\downarrow\downarrow}_{L,n}+2q+1$ & $v^{\uparrow\downarrow\downarrow}_{L,n}+2q+2$ &
$\Delta Z_{\uparrow\downarrow\downarrow\, \rm triplon}^{\rm rel}=\Delta Z_{1,2}^{\rm rel}-Z_1\Delta B_{1,1}=\Delta B_{1,2}$ \\
\hline
\end{tabular}
\end{center}
\caption{For $3+1$ or $2+2$ trapped fermions, the relevant {\sl decoupled asymptotic objects} (DAOs), their {\sl internal} energy levels $\epsilon_{\rm int}^{\rm unit}$ and $\epsilon_{\rm int}^{\rm ideal}$ and the difference (indicated by the letter $\Delta$) of their {\sl relative} partition functions in the unitary and ideal gas cases. Each DAO moves as a free particle in the trap, and {\sl internal} or {\sl relative} means after removal of this centre-of-mass motion.  All integers $q$, $L$, $n$ run over $\mathbb{N}$.  The pairon internal states, which correspond to the interaction-sensitive states of the relative motion of two opposite-spin fermions, have a zero angular momentum and are not degenerate.  The triplon internal energies of angular momentum $L$ are related to the positive roots $u_{L,n}$ or poles $v_{L,n}$ of Efimov's three-body transcendental function $s\mapsto \Lambda^{\uparrow\uparrow\downarrow}_L(s)$ or $s\mapsto \Lambda^{\uparrow\downarrow\downarrow}_L(s)$, and have a degeneracy $2L+1$.  The effective partition function of a triplon involves a subtraction of the partition function of two associated emerging DAOs (one atom and one pairon) to avoid double-counting.  The equation (\ref{eq:DB11}) was used, as well as the identity (\ref{eq:DB21}) and its $\uparrow \leftrightarrow \downarrow$ counterpart.  
\label{table:DAO}}
\end{table}
Considering the absence of atom-triplon interaction, one has
\be
\Delta Z_{\uparrow\,\mathrm{atom}+\uparrow\uparrow\downarrow\,\mathrm{triplon}}^{\rm rel}= Z_1 \Delta B_{2,1}
\label{eq:dzattrip_seconde}
\ee
where the factor $Z_1$ is the partition function of the atom-triplon relative motion in the trap.  When combined with (\ref{eq:DB31red}) and (\ref{eq:DBconj})  this shows that the new conjecture coincides with the old one for the $3+1$ system:
\be
\Delta B_{3,1}^{\rm new\, conj}=\Delta B_{3,1}^{\rm old\, conj}
\label{eq:newconj31}
\ee
This is partly accidental as the old conjecture was a guess.  Two inspiring rewritings of the above equations will be invaluable in what follows. First, we rewrite the conjecture~(\ref{eq:C31}) in a {\sl mathematically} insightful way, that better reveals the key role played by the continuous spectrum $k\mapsto\Lambda^{\uparrow\uparrow\downarrow}_L(\ii k)$, $k\in\mathbb{R}$ and $L\in \mathbb{N}$, in the failure of the na{\"\i}ve residue formula. Using (\ref{eq:DB21ex}) and (\ref{eq:dzattrip_seconde}) we get 
\be
I_{3,1}=\frac{S_{3,1}^{\rm new\, conj}}{2\sinh\bar{\omega}} -Z_1 \sum_{L\in\mathbb{N}} (2L+1) \int_\mathbb{R} \frac{\dd S}{2\pi} \frac{\sin(\bar{\omega}S)}{2\sinh\bar{\omega}} \frac{\dd}{\dd S} \left[ \ln \Lambda^{\uparrow\uparrow\downarrow}_L(\ii S)\right] \label{eq:insight1}
\ee
where the factor $Z_1$ originates from the partition function of the DAOs relative motion.  Second, we attribute the coincidence of the old and new conjectures to the absence of atom-triplon correlations, hence the {\sl physically} insightful rewriting of equation~(\ref{eq:newconj31}):
\be
\Delta B_{3,1}^{\rm new\, conj}=\Delta B_{3,1}^{\rm old\, conj}+ Z_1^{-1} (\Delta Z_{\uparrow\,\mathrm{atom}+\uparrow\uparrow\downarrow\,\mathrm{triplon}}
-Z_{\uparrow\,\mathrm{atom}} \Delta Z_{\uparrow\uparrow\downarrow\,\mathrm{triplon}})
\label{eq:insight2}
\ee
 
The $2+2$ fermion case is richer. It leads to the expected continuous spectrum branches $k\mapsto \Lambda_L^{\uparrow\uparrow\downarrow}(\ii k)$ and $k\mapsto \Lambda_L^{\uparrow\downarrow\downarrow}(\ii k)$ of degeneracies $2L+1$, $k\in\mathbb{R}$ and $L\in\mathbb{N}$ \cite{pasEfim4}, associated to the asymptotic splitting of the four fermions into one $\downarrow$ fermion plus one $\uparrow\uparrow\downarrow$ triplon, and one $\uparrow$ fermion plus one $\uparrow\downarrow\downarrow$ triplon.  But the continuous spectrum of $\mathcal{M}_\ell^{(\sigma)}(\ii S)$ for $\sigma=+1$ has an additional, nondegenerate branch $k\mapsto \frac{1}{\sqrt{2}}[1-\frac{(-1)^\ell}{\cosh (k\pi/2)}]$, corresponding to two decoupled {\sl pairons}, that is $s$-wave correlated pairs of neighbouring $\uparrow\downarrow$ particles, with relative orbital angular momentum $\ell$ \cite{pasEfim4}.  A pairon is a freely moving bosonic particle in the trap, with an internal structure given in the table \ref{table:DAO} that reproduces the asymptotic three-body spectrum\footnote{After removal of the centre of mass the energy levels of a pairon and a $\uparrow$ particle in the trap are $(2q+\frac{1}{2}+2n+\ell+\frac{3}{2})\hbar\omega$ in the unitary limit.  In the limit of large quantum numbers $\ell\gg 1$ or $n\gg 1$ this must differ from $(u_{\ell,n}^{\uparrow\uparrow\downarrow}+2q+1)\hbar\omega$ by $o(1)\hbar\omega$, which is confirmed by the exact asymptotic analysis of reference \cite{Werner3corps}. For the ideal gas there is an additional term $\hbar \omega$ and one recovers the result~(\ref{eq:s2p1}).}.  The {\sl mathematical} formulation (\ref{eq:insight1}) of the new conjecture immediately becomes
\bea
I_{2,2}=\frac{S_{2,2}^{\rm new\, conj}}{2\sinh\bar{\omega}}  \nonumber \\
-Z_1 \sum_{L\in\mathbb{N}} (2L+1) \int_\mathbb{R} \frac{\dd S}{2\pi} \frac{\sin(\bar{\omega}S)}{2\sinh\bar{\omega}} \frac{\dd}{\dd S} \left[
\ln \Lambda^{\uparrow\uparrow\downarrow}_L(\ii S)+\ln \Lambda^{\uparrow\downarrow\downarrow}_L(\ii S)\right] \\
-\sum_{n,\ell} (2\ell+1) e^{-\bar{\omega}(2n+\ell+\frac{3}{2})}  
\int_\mathbb{R} \frac{\dd S}{2\pi} \frac{\sin(\bar{\omega}S)}{2\sinh\bar{\omega}} \frac{\dd}{\dd S} 
\ln\left[1-\frac{(-1)^\ell}{\cosh\frac{S\pi}{2}}\right]
\eea
Due to the identity (\ref{eq:DB21ex}) and its $1+2$ counterpart, the second term reduces to $-Z_1 (\Delta B_{2,1}+\Delta B_{1,2})$, which partially reconstructs the right-hand side of equation~(\ref{eq:DB22red}).  In the third term, it is apparent that $Z_1$ can no longer be factored out and that one must keep a sum over the quantum numbers $n$ and $\ell$ of the pairons relative motion, since the additional branch of the continuous spectrum depends on $\ell$; the integral over $S$ can be calculated exactly by taking the sum and the difference of the odd and even $\ell$ integrals  and using $\int_{\mathbb{R}} \dd S \frac{\sin(xS)}{\sinh(S\pi)}=\tanh\frac{x}{2}, x\in \mathbb{R}$; the result differs from the last term $-Z_1 \Delta B_{1,1}^2/2$ of equation~(\ref{eq:DB22red}). The new conjecture for the $2+2$ system thus differs from the old one:
\be
\Delta B_{2,2}^{\rm new\, conj}= \Delta B_{2,2}^{\rm old\, conj} + \frac{1}{32} \frac{1}{\cosh^3\frac{\bar{\omega}}{2}\cosh\bar{\omega}}
\label{eq:newconj22}
\ee
To obtain the {\sl physical} formulation (\ref{eq:insight2}) of the new conjecture, one must realize that, contrarily to the distinguishable atom and triplon, the pairons are identical bosons, which induces statistical correlations among them even if they do not interact, so
\be
\Delta B_{2,2}^{\rm new\, conj}=\Delta B_{2,2}^{\rm old\, conj} +Z_1^{-1} \left[\Delta Z_{\rm 2\, pairons}-\frac{1}{2} \Delta(Z_{\rm 1\, pairon}^2)\right]
\label{eq:forme2pour22}
\ee
where $\Delta$ still represents the difference between the unitary gas and ideal gas values, for example $\Delta(Z_{\rm 1\, pairon}^2)=(Z_{\rm 1\, pairon}^{\rm unit})^2-(Z_{\rm 1\, pairon}^{\rm ideal})^2$.  In the sum over the internal quantum numbers $q$ and $q'$ of each pairon appearing in $Z_{\rm 2\, pairons}$, the internal states $(q,q')$ and $(q',q)$ are physically equivalent and shall not be double-counted.  Also, when the relative angular momentum $\ell$ of the pairons is odd, their internal state must be antisymmetric, which excludes the state $(q,q)$. In the unitary gas case, this leads to the relative two-pairon partition function
\bea
Z_{\rm 2\, pairons}^{\rm rel, unit} &=& \!\!\sum_{n,\ell}^{\ell\,\mathrm{even}} (2\ell+1) e^{-\bar{\omega}(2n+\ell+\frac{3}{2})}
\left[\sum_{q} e^{-2\bar{\omega}(2q+\frac{1}{2})} + \frac{1}{2}\sum_{q\neq q'} e^{-\bar{\omega}(2q+\frac{1}{2}+2q'+\frac{1}{2})}\right] \nonumber\\
&+& \!\sum_{n,\ell}^{\ell\,\mathrm{odd}} (2\ell+1) e^{-\bar{\omega}(2n+\ell+\frac{3}{2})} \left[ \frac{1}{2}\sum_{q\neq q'} e^{-\bar{\omega}(2q+\frac{1}{2}+2q'+\frac{1}{2})}\right]
\label{eq:Z2pairons}
\eea
We replace each term $\frac{1}{2}$ under the exponentials by $\frac{3}{2}$ to obtain $Z_{\rm 2\, pairons}^{\rm rel, ideal}$ and form the difference $\Delta Z_{\rm 2\, pairons}^{\rm rel}=Z_{\rm 2\, pairons}^{\rm rel, unit}-Z_{\rm 2\, pairons}^{\rm rel, ideal}$.  This leads in equation~(\ref{eq:forme2pour22}) to exactly the same result as in equation~(\ref{eq:newconj22}), which was not granted and is a good consistency check.

To be complete, and make the new conjecture as physically transparent as possible, we note that it takes {\sl a posteriori} a very simple form if, rather than using the four-body sums $S_{N_\uparrow,N_\downarrow}$ or cluster coefficients $\Delta B_{N_\uparrow,N_\downarrow}$, one turns back to the four-body partition functions $\Delta Z_{N_\uparrow,N_\downarrow}^{\rm rel}$ of equations (\ref{eq:DB31},\ref{eq:DB22}):
\bea
\label{eq:transpa31}
I_{3,1} &=& \Delta Z_{3,1}^{\rm rel, new\, conj}-Z_{2,0}\Delta B_{1,1} - Z_1 \Delta B_{2,1} \\
\label{eq:transpa22}
I_{2,2} &=& \Delta Z_{2,2}^{\rm rel, new\, conj} - Z_1 \Delta B_{2,1} - Z_1 \Delta B_{1,2} 
-\Delta Z_{\rm 2\, pairons}^{\rm rel}  \nonumber \\
&&- Z_1 (Z_1-Z_{\rm 1\, pairon}^{\rm rel, ideal}) \Delta B_{1,1}
\eea
In the equation (\ref{eq:transpa31}) the second and third terms in the right-hand side correspond to the two possible splittings of $\uparrow\uparrow\uparrow\downarrow$ into DAOs, respectively $(\uparrow) + (\uparrow) + (\uparrow\downarrow)$ and $(\uparrow) + (\uparrow \uparrow\downarrow)$; double-counting is avoided by the way the triplon partition function is calculated, see the table \ref{table:DAO}. In the equation (\ref{eq:transpa22}), the second, third, fourth and fifth terms in the right-hand side correspond to the four possible splittings of $\uparrow\uparrow\downarrow\downarrow$ into DAOs, respectively $(\downarrow) + (\uparrow \uparrow\downarrow)$, $(\uparrow) + (\uparrow\downarrow\downarrow)$, $(\uparrow\downarrow) + (\uparrow\downarrow)$ and $(\uparrow) + (\downarrow) + (\uparrow\downarrow)$. The only subtlety lies in the fifth term: one must include only the interaction-insensitive states of the $(\uparrow) + (\downarrow)$ system, hence the subtraction of $Z_{\rm 1\, pairon}^{\rm rel, ideal}$ from the partition function $Z_1$ of their relative motion, to avoid a double-counting with the first pairon contribution of the fourth term.

The new conjectured value (\ref{eq:newconj22}) for $\Delta B_{2,2}$ corresponds to the upper black solid line in the right panel of figure~\ref{fig:blvscj}. It is now in good agreement with the numerics.  Its $\bar{\omega}\to 0^+$ limit $\Delta B^{\rm new\, conj}_{2,2}(0^+)= -0.0305(2)$ is of the same order as the approximate value $\Delta B_{2,2}(0^+)=-0.036$ of reference \cite{Levinsen}.  There is therefore a good possibility that the new conjecture for $\Delta B_{2,2}$ is exact.

\begin{figure}[htb]
\begin{center}
\includegraphics[width=10cm,clip=]{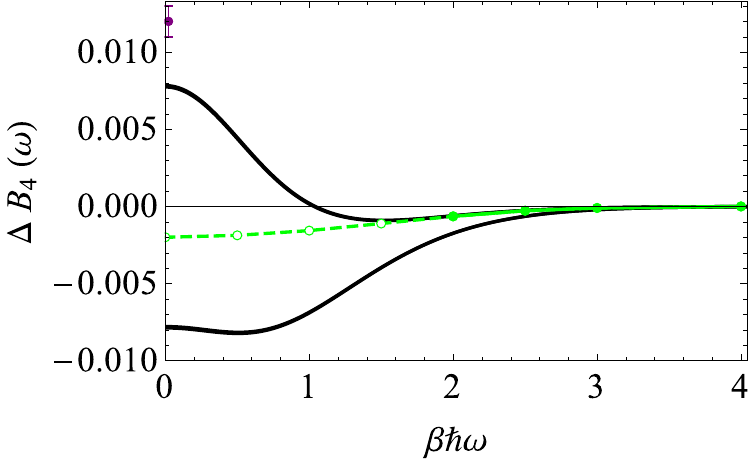}
\end{center}
\caption{Total fourth-order cluster coefficient $\Delta B_4(\omega)$ of the trapped unpolarised spin-$1/2$ unitary Fermi gas, as a function of $\beta\hbar\omega$ as in figure~\ref{fig:blvscj}. Upper (lower) black solid line: the new (old) conjecture. Green line with symbols: numerical results of reference~\cite{Blumeb4} (disks connected by a solid line: actually calculated values; circles connected by a dashed line: values resulting from an extrapolation).  Symbol with an error bar: most recent experimental result \cite{virielMIT}.}
\label{fig:la_somme}
\end{figure}

In figure~\ref{fig:la_somme} we plot the total fourth-order cluster coefficient of the trapped unpolarised spin-1/2 unitary Fermi gas, or more precisely its deviation $\Delta B_4(\omega)=\frac{1}{2} [\Delta B_{3,1}(\omega)+\Delta B_{2,2}(\omega)+\Delta B_{1,3}(\omega)]$ from the ideal gas value, as a function of $\hbar\omega/k_B T$. The new (old) conjecture corresponds to the upper (lower) black solid line. Remarkably, it is not monotonic. Its $\omega\to 0^+$ limit is related to the homogeneous gas value $\Delta b_4$ by $\Delta B_4(0^+)=\frac{1}{8}\Delta b_4$, where the homogeneous gas cluster expansion for the pressure difference $\Delta P=P_{\rm unit}-P_{\rm ideal}$ takes the form $\Delta P\lambda_{\rm dB}^3/k_B T=2\sum_{n\geq 1} \Delta b_n z^n$ with $z_\uparrow=z_\downarrow=z$ and $\lambda_{\rm dB}=(2\pi\hbar^2/m k_B T)^{1/2}$.  It can be compared to the most recent experimental result $\Delta b_4=0.096(10)$ \cite{virielMIT}: while the old conjecture had even its sign wrong, the new conjecture, leading to $\Delta B_4^{\rm new\, conj}(0^+)=0.00775(10)$ and $\Delta b_4^{\rm new\, conj}=0.062(1)$, is not far but still off by more than two standard deviations. We conclude that either the new conjecture is wrong, or there is an underestimated systematic error in the experimental result due to the extrapolation to $z=0$ of data all having for accuracy reasons a fugacity $z>1$ (see the figure 4.11 in reference \cite{TheseNascimbene}). The recent path integral Monte Carlo result $\Delta b_4^{\rm PIMC}(0^+)=0.078(18)$ \cite{BlumePIMC}
is almost exactly halfway and does not allow to arbitrate.

\begin{figure}[htb]
\begin{center}
\includegraphics[width=10cm,clip=]{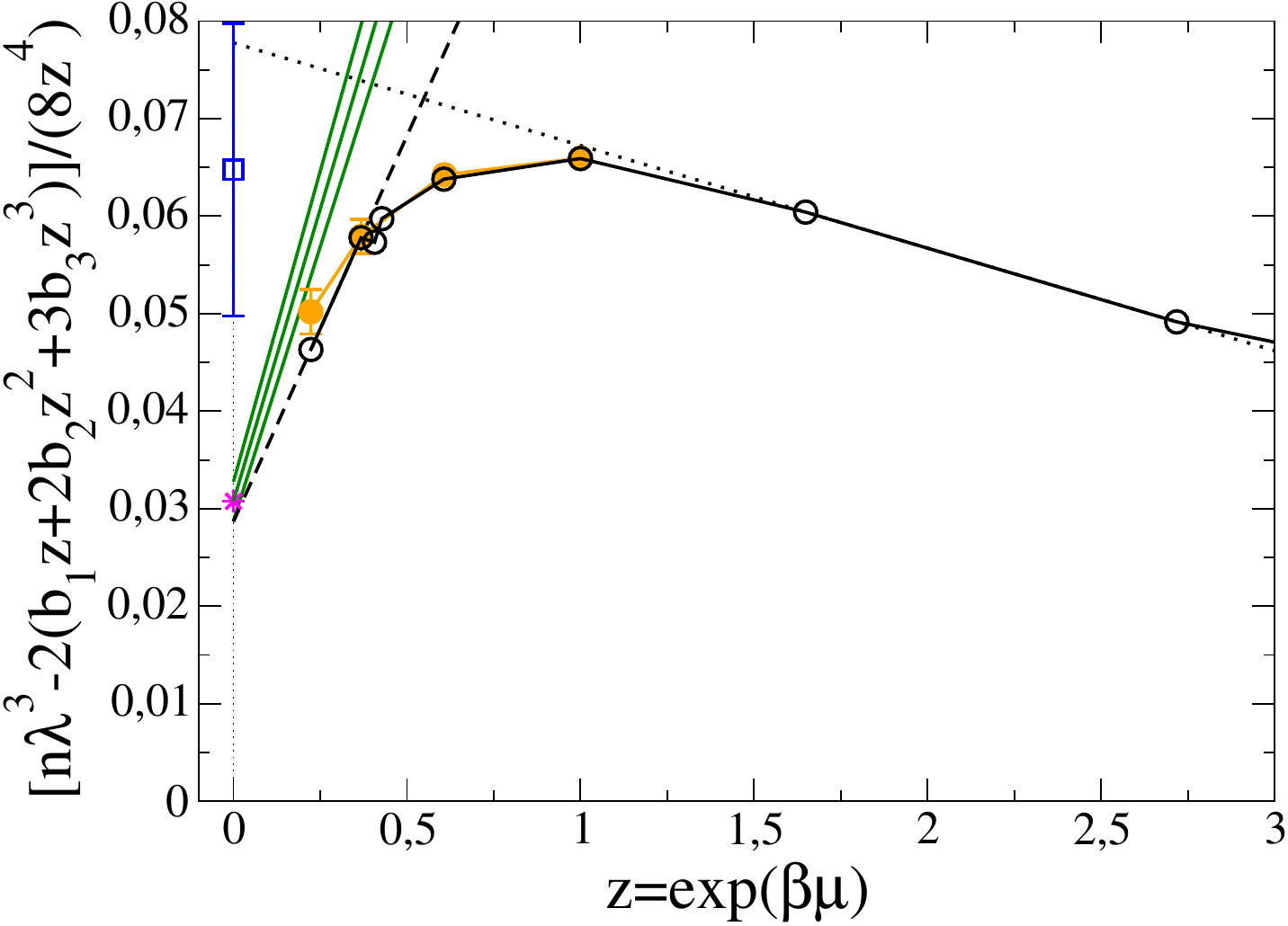}
\end{center}
\caption{(figure added to the published version) Circles: diagrammatic Monte Carlo results of reference [Nature Physics {\bf 8}, 366 (2012)] for the unpolarised uniform unitary Fermi gas, as functions of the fugacity $z$. $n$ is the total density of the gas and $\lambda$ is the thermal de Broglie wavelength. From $n\lambda^3$ one subtracts the third order cluster expansion so that the resulting function, when divided by $8z^4$, tends to the fourth cluster coefficient $b_4$ when $z$ tends to 0.  Dotted line: linear extrapolation of the right-most two points to $z=0$, giving a value of $b_4$ within the error bars of the ENS experimental value \cite{virielENS} (blue square with error bars).  Dashed line: linear extrapolation of the left-most two points to $z=0$, giving a value of $b_4$ very close to our new conjecture (magenta star).  Orange disks with error bars: diagrammatic Monte Carlo results of reference [Phys. Rev. Lett. {\bf 121}, 130405 (2018)]. Green solid lines: affine approximation $b_4+(5/4) b_5 z$ obtained by replacing $n\lambda^3$ by its fifth-order cluster expansion and using the values of $b_4$ and $b_5$ of reference [Phys. Rev. Lett. {\bf 125}, 050403 (2020)]; the upper (lower) line corresponds to the upper (lower) bounds of $b_4$ and $b_5$ uncertainty intervals, and the middle line corresponds to the midpoints of the uncertainty intervals.}
\label{fig:MCdiag}
\end{figure}

{\bf Note added to the published version:} A precise calculation of the equation of state of the unpolarised unitary Fermi gas was performed with the diagrammatic Monte Carlo technique, see [K. Van Houcke, F. Werner, E. Kozik, N. Prokof'ev, B. Svistunov, M.J.H. Ku, A.T. Sommer, L.W. Cheuk, A. Schirotzek, and M.W. Zwierlein, Nature Physics {\bf 8}, 366 (2012)]. As shown in figure \ref{fig:MCdiag}, the Monte Carlo data point to a value of the fourth cluster coefficient $b_4$ of the uniform unitary Fermi gas in agreement with our new conjecture (we recall that $b_n=\Delta b_n+\frac{(-1)^{n+1}}{n^{5/2}}$ so that $b_4=\Delta b_4-\frac{1}{32}$). They also explain why a higher value of $b_4$ is obtained if one extrapolates the data from an interval of fugacity $z>1$ as done with the ENS experimental results in references \cite{virielENS,TheseNascimbene}. We hope that this will stimulate more precise experimental measurements, not only in the unpolarised case (where only $b_4$ can be accessed) but also in the weakly spin-polarised case where the spin susceptibility gives access to the cluster coefficients $b_{3,1}$ and $b_{2,2}$ separately. {\bf Update of added note:} A refined diagrammatic Monte Carlo calculation of the equation of state of the unpolarised unitary Fermi gas was performed in [R. Rossi, T. Ohgoe, K. Van Houcke, F. Werner, Phys. Rev. Lett. {\bf 121}, 130405 (2018)], essentially confirming the previous diagrammatic Monte Carlo results. Also, an accurate numerical calculation of $b_4$ and $b_5$ was performed in [Y. Hou, J.E. Drut, Phys. Rev. Lett. {\bf 125}, 050403 (2020)], giving $\Delta b_4=0.062(2)$ and $\Delta b_5=0.078(6)$; the obtained value of $\Delta b_4$ perfectly agrees with our new conjecture. We have correspondingly updated figure \ref{fig:MCdiag}.

\section{Conclusion}
\label{sec:conclusion}

We have determined the interaction-sensitive states of a harmonically trapped two-component ideal Fermi gas, using a Faddeev ansatz for the $N$-body wavefunction.

We have found a simple rule to obtain the interaction-sensitive relative or internal energy levels as follows.  One removes one spin $\uparrow$ fermion and one spin $\downarrow$ fermion to build a pair of particles with a zero relative orbital momentum, which renders the state interaction-sensitive. One freely distributes the remaining $N_\uparrow-1$ spin $\uparrow$ fermions and the remaining $N_\downarrow-1$ spin $\downarrow$ fermions among the single-particle energy levels of the harmonic oscillator, in a way compatible with Fermi statistics. One adds to the resulting energy levels the energy $(2q+\frac{3}{2})\hbar\omega$, where $\frac{3}{2}\hbar\omega$ may be interpreted as the internal energy of the subtracted $\uparrow\downarrow$ pair and $2q\hbar\omega$, with $q$ running over all natural integers, corresponds to the quantised excitation spectrum of the collective breathing mode of our SO(2,1)-symmetric system.

This simple rule must be refined because some of its energy levels actually correspond to a zero Faddeev ansatz for the wavefunction, by destructive interference of individually nonzero Faddeev components, and are unphysical solutions. This problem was known for $2+1$ fermions, where there is a single unphysical solution.  We studied it for $3+1$ and $2+2$ fermions and we found a countable infinite number of unphysical solutions. A Fourier space reasoning leads to a class of unphysical solutions that lends itself to a simple picture: everything happens as if each particle of the subtracted $\uparrow\downarrow$ pair was prepared in the harmonic oscillator ground state $|0,0,0\rangle$; the unphysical solutions in question then formally correspond to putting one of the $N_\uparrow-1$ remaining spin $\uparrow$ particles or one of the $N_\downarrow-1$ remaining spin $\downarrow$ particles in the state $|0,0,0\rangle$, thus ``violating" the Pauli exclusion principle.  In the case of $2+2$ fermions, however, this is not the end of the story: there exist additional unphysical solutions, as a comparison to a numerically calculated four-body spectrum shows. We could find them on a case by case basis by a real space calculation, with linear algebra within a set of homogeneous polynomials of fixed total angular momentum, degree and parity.

We have applied the above results to the cluster or virial expansion of a two-component unitary Fermi gas. To evaluate the cluster coefficients with the harmonic regulator technique, one needs to calculate the difference of the partition functions of the interaction-sensitive energy levels of the unitary gas and of the ideal gas, which makes the link with our problem.  For an arbitrary particle number, we have defined a generalised transcendental Efimov function $\Lambda(s)$ such that the interaction-sensitive energy levels for the unitary gas can be expressed in terms of the positive roots $u_n$ of $\Lambda(s)$.  We then showed that the interaction-sensitive energy levels of the ideal gas can be expressed in terms of the positive poles $v_n$ of $\Lambda(s)$.  We reached an optimised writing of the third and fourth cluster coefficients in terms of the sums $\sum_n (e^{-\bar{\omega}u_n}-e^{-\bar{\omega}v_n})$, with $\bar{\omega}=\hbar\omega/(k_B T)$, which evoke the residues of Cauchy's theorem applied to a contour integration of $s\mapsto e^{-s\bar{\omega}} \frac{\dd}{{\dd} s} \ln\Lambda(s)$.  This optimised writing allowed us to extend the applicability domain of the numerical calculations of the cluster coefficients $\Delta B_{3,1}$ and $\Delta B_{2,2}$ of the reference \cite{Blumeb4} to lower values of $\bar{\omega}$, before they diverge due to the energy cut-off in the numerics.  Over this range of values of $\bar{\omega}$, $1\lesssim \bar{\omega}$, it shows that the blind conjecture given in the reference \cite{pasEfim4} for $\Delta B_{3,1}$ is accurate, while the one for $\Delta B_{2,2}$ disagrees by a factor $\simeq 2$. Using a physical reasoning, we have constructed a new conjecture in terms of the decoupled asymptotic objects (DAOs) emerging in the four-body interaction-sensitive spectrum at large excitation amplitudes or quantum numbers, i.e.\ individual atoms, pairons and triplons, in the unitary and ideal gases.  The new conjecture gives the same value for $\Delta B_{3,1}$, and a new, now accurate value for $\Delta B_{2,2}$. The failure of the old conjecture for $\Delta B_{2,2}$ results {\sl a posteriori} from the omission of the statistical correlations between pairons induced by their bosonic nature.  The most recent numerical calculations, based on a dedicated path integral Monte Carlo approach, lead to a $\bar{\omega}\to 0$ limit in agreement with our new conjecture within error bars \cite{BlumePIMC}, but are not (yet) accurate enough to exclude the possibility that this agreement is fortuitous.

\section*{Acknowledgements}
S.E.\ thanks the Japan Society for the Promotion of Science (JSPS) for support. The group of Y.C.\ is affiliated to IFRAF.

\end{document}